\documentclass[twocolumn,showpacs,amsmath,amssymb,superscriptaddress]{revtex4-1}
\usepackage{dcolumn}
\usepackage{color}
\usepackage{graphicx}
\usepackage{amsmath}
\usepackage{multirow}
\usepackage{bm}
\usepackage{url}

\begin{document}

\title{Structure of krypton isotopes within
the interacting boson model derived from the Gogny energy density functional}

\author{K.~Nomura}
\affiliation{Physics Department, Faculty of Science, University of
Zagreb, HR-10000 Zagreb, Croatia}
\affiliation{Center for Computational
Sciences, University of Tsukuba, Tsukuba 305-8577, Japan}

\author{R.~Rodr\'iguez-Guzm\'an}
\affiliation{Physics Department, Kuwait University, 13060 Kuwait, Kuwait}

\author{Y.~M.~Humadi}
\affiliation{Physics Department, Kuwait University, 13060 Kuwait, Kuwait}

\author{L.~M.~Robledo}
\affiliation{Departamento de F\'\i sica Te\'orica, Universidad
Aut\'onoma de Madrid, E-28049 Madrid, Spain}

\author{H.~Abusara}
\affiliation{Physics Department, Birzeit University, Birzeit, Palestine}

\date{\today}

\begin{abstract}

The evolution and coexistence of the nuclear shapes as well as the 
corresponding low-lying collective states and electromagnetic transition rates
are investigated along the Krypton isotopic chain within the 
framework of the interacting boson model (IBM). The IBM 
Hamiltonian is determined through mean-field calculations based on 
the several parametrizations of the Gogny energy density functional 
and the relativistic 
mean-field Lagrangian. The mean-field energy surfaces, as functions 
of the axial $\beta$ and triaxial $\gamma$ quadrupole deformations, are 
mapped onto the expectation
 value of the  interacting-boson Hamiltonian that explicitly
 includes the particle-hole excitations. The  resulting 
 boson Hamiltonian is
 then  used to compute low-energy excitation
 spectra as well as E2 and E0 transition probabilities for
 $^{70-100}$Kr. Our results point to a number of examples of the
 prolate-oblate shape transitions and coexistence both on the
 neutron-deficient  
and neutron-rich sides. A reasonable agreement with the available 
experimental data is obtained for the considered nuclear properties. 

\end{abstract}

\keywords{}

\maketitle


\section{Introduction}


The low-lying structure of Kr isotopes is characterized by a rich 
variety of shape phenomena like shape transitions when neutron number 
is varied \cite{CasBook,cejnar2010} as well as shape coexistence and 
mixing \cite{heyde2011}. On the neutron-deficient side, especially in 
the case of isotopes  with approximately the same number of  protons 
and neutrons, the experimental evidence \cite{clement2007} regarding 
the emergence of prolate-oblate shape coexistence and mixing has 
already been studied using different theoretical frameworks 
\cite{bender2006,clement2007,fu2013,trodriguez2014,PETROVICI2000,SATO2011}. Low-lying excited 
$0^+$ states have been observed for some Kr nuclei, e.g., $^{72-78}$Kr. 
Those states have been associated to  intruder excitations 
\cite{heyde2011}.

In the last few years, it has become possible to access neutron-rich Kr 
nuclei experimentally 
\cite{naimi2010,albers2012,albers2013,rzkacaurban2017,dudouet2017,flavigny2017}. 
Even those isotopes beyond the neutron number $N\approx 60$ have been 
experimentally studied, as reported quite recently in 
Refs.~\cite{dudouet2017,flavigny2017} where  the spectroscopy of the 
radioactive isotopes $^{96,98,100}$Kr is analyzed. The structural 
evolution in  neutron-rich nuclei with mass number $A\approx 100$ is 
rather sensitive to the underlying shell structure and such 
experimental information is quite useful to deepen our understanding of 
it and offers the possibility to learn about unique features related to 
shape transitions in neutron-rich Kr isotopes. For instance, in 
contrast to its neighboring neutron-rich Sr and Zr nuclei where the 
shape transition is suggested to take place rather rapidly around 
$N=60$ \cite{togashi2016,kremer2016,clement2017}, the onset of 
deformation is shown to emerge much more moderately  along the Kr 
isotopic chain \cite{albers2012,albers2013,dudouet2017}.

From a theoretical point of view,  the large-scale shell model 
\cite{caurier2005} and the nuclear energy density functional (EDF) 
\cite{bender2003} approaches are among the most popular microscopic 
nuclear structure models for medium-heavy and heavy nuclei. The former 
allows direct access to the spectroscopic properties via the 
diagonalization of the Hamiltonian matrix defined in the corresponding 
configuration space. However, in open-shell regions with increasing 
number of valence nucleons, the dimension of the shell-model matrix 
becomes exceedingly large making a systematic investigation of the 
nuclear spectroscopy less tractable. On the other hand, the EDF 
framework allows the systematic study of several nuclear properties all 
over the nuclear chart. A number of self-consistent mean-field (SCMF) 
calculations with both non-relativistic \cite{bender2003} and 
relativistic \cite{vretenar2005,niksic2011} EDFs have so far been 
performed to investigate structural phenomena in atomic  nuclei. 
Nevertheless, a more quantitative analysis of shape transitions 
requires the extension of the mean-field framework so as to include 
beyond-mean-field correlations associated with the restoration of 
broken symmetries and/or fluctuations in the collective parameters 
within the symmetry-projected Generator Coordinate Method (see, for 
example, \cite{rayner2002,bender2003,niksic2011}). Though quite robust, 
the method becomes increasingly difficult to implement from a 
computational point of view in the case of heavy nuclei and/or when 
several collective coordinates have to be considered in the 
symmetry-projected GCM ansatz.

To alleviate the computational effort required in symmetry-projected 
GCM configuration mixing calculations several approximations have 
already been employed. Among them we mention here the five-dimensional 
collective Hamiltonian (5DCH) approach, based on both non-relativistic 
\cite{delaroche2010} and relativistic \cite{niksic2011} EDFs, and the 
fermion-to-boson mapping procedure that allows to build an algebraic 
model of interacting bosons \cite{nomura2008} starting from a given 
EDF. In this study, we resort to the latter approach \cite{nomura2008} 
and use the microscopic energy surface of a given nucleus, obtained via 
constrained mean-field calculations \cite{RS}, as input to a mapping 
procedure involving the intrinsic wave function of the boson system 
taking into account  particle-hole excitations. This mapping procedure 
allows the determination of the parameters of the corresponding 
(bosonic) IBM Hamiltonian that is subsequently used to compute the 
excitation spectra and electromagnetic transition rates for a give 
nuclear system. At variance with the phenomenological  IBM 
calculations, where the parameters of the Hamiltonian are fitted to 
reproduce spectroscopic data, within the  already mentioned 
fermion-to-boson mapping procedure \cite{nomura2008} the corresponding 
IBM Hamiltonian is obtained from microscopic EDF calculations and, 
therefore, the method can be extrapolated to regions of the nuclear 
chart where experimental data are scarce or even not available. Several 
applications of the fermion-to-boson mapping procedure have been 
reported in the literature. For instance, it has recently been employed 
to describe shape transitions and shape coexistence in Ru, Mo, Zr and 
Sr isotopes with mass number $A\approx 100$ \cite{nomura2016zr} as well 
as for neutron-rich Ge and Se nuclei with $70\leq A\leq 90$ 
\cite{nomura2017ge}.

In this study, we have resorted to the parametrization D1M \cite{D1M} 
of the Gogny EDF to obtain, via mean-field calculations, the required 
microscopic input used to build the IBM Hamiltonian. To examine the 
robustness of our fermion-to-boson mapping procedure with respect to 
the underlying EDF, calculations have  been performed with two other 
parametrizations of the Gogny EDF, i.e., D1S \cite{D1S} and D1N 
\cite{D1N}. Furthermore, mean-field calculations have also been carried 
out with the density-dependent meson-exchange (DD-ME2) 
\cite{lalazissis2005} and point-coupling (DD-PC1) \cite{DDPC1} 
relativistic EDFs. Nevertheless, in our discussions we will mainly 
focus on the results obtained with the Gogny-D1M EDF since, as will be 
shown, the results to be presented later on in this study do not depend 
significantly on the underlying EDF employed in the mapping procedure.

The paper is organized as follows. In Sec.~\ref{sec:model}, we briefly 
outline the fermion-to-boson mapping procedure employed in this work to 
study the isotopes $^{70-100}$Kr. The results of our calculations are 
presented in Sec.~\ref{sec:results}. First, in Sec.~\ref{sec:pes}, we 
discuss the  microscopic energy surfaces obtained at the mean-field 
level as well as the mapped IBM ones. The IBM parameters derived via 
the mapping procedure and the configurations employed in the 
calculations are presented in Sec.~\ref{sec:config}. In 
Secs.~\ref{sec:energy} and \ref{sec:transition}, we turn our attention 
to spectroscopic properties such as the systematics of the energy 
spectra and the transition rates predicted in our calculations as well 
as to the comparison with the available experimental data. The detailed 
spectroscopy of a selected sample of Kr isotopes is discussed in 
Sec.~\ref{sec:detail}. In Sec.~\ref{sec:uncertainty}, we consider the 
sensitivity of the results with respect to the underlying EDF used in 
the mapping procedure. Finally, Sec.~\ref{sec:summary} is devoted to 
the concluding remarks and work perspectives.


\section{Description of the model\label{sec:model}}


In this section, we briefly outline the fermion-to-boson mapping
procedure employed in this work. For a more detailed account, the
reader is referred to \cite{nomura2016zr,nomura2017ge} and references  
therein.


\subsection{SCMF calculations}


The first step in our procedure is to perform a set of constrained SCMF 
calculations, within the Hartree-Fock-Bogoliubov (HFB) method and based 
on the Gogny-D1M EDF \cite{rayner2010pt}. We have also carried out 
mean-field calculations with the  parametrizations D1S \cite{D1S} and 
D1N \cite{D1N}  of the Gogny-EDF as well as with the relativistic 
DD-ME2 \cite{lalazissis2005} and DD-PC1 \cite{DDPC1} EDFs. In this way 
we obtain the HFB deformation energy surfaces parametrized by the usual 
quadrupole shape degrees of freedom $\beta$ and $\gamma$ \cite{BM}. 
Here, we have used constraints on the operators $\hat Q_{20}$ and $\hat 
Q_{22}$. They are related to the deformation parameters $\beta$ and 
$\gamma$ through the relations $\beta=\sqrt{4\pi/5}\sqrt{\langle\hat 
Q_{20}\rangle^2+\langle\hat Q_{22}\rangle^2}/\langle r^2\rangle$ and 
$\gamma=\arctan{(\langle\hat Q_{22}\rangle/\langle\hat 
Q_{20}\rangle)}$, respectively. In the previous expressions, $\langle 
r^2\rangle$ denotes the mean-square radius obtained from the 
corresponding  HFB state.

\subsection{IBM framework}
\label{IBM-Hamilt-Rayner}

The building blocks of the IBM system, that predominantly determine the 
low-energy quadrupole collective states, are the $J^{\pi}=0^+$ ($s$) 
and $2^+$ ($d$) bosons which represent the collective $J^{\pi}=0^+$ and 
$2^+$ pairs of valence nucleons, respectively \cite{IBM,OAI}. 
Therefore, the number of bosons $n_b$ equals that of pairs of valence 
nucleons (particle or hole) \cite{IBM,OAI}.  The boson Hamiltonian is 
diagonalized in a given valence space (one major shell). In the present 
work, we use the same model space for the boson system as in our 
previous study \cite{nomura2017ge}, i.e, the proton $Z=28-50$ major 
shell and the neutron $N=28-50$ (for $^{70-86}$Kr) and $N=50-82$ (for 
$^{88-100}$Kr) major shells. For the sake of simplicity, no distinction 
is made  between proton and neutron bosons.

For many of the studied Kr isotopes, the Gogny-D1M HFB energy surfaces 
exhibit more than one minimum, reflecting a pronounced competition 
between different intrinsic configurations. Previous IBM calculations  
already suggest that the  low-lying $0^+_2$ state in neutron-deficient 
Kr isotopes, as well as in the neighboring Se and Ge nuclei, could 
arise from particle-hole excitations and therefore have an intrinsic 
structure different from the one of the ground state 
\cite{kaup1979,duval1983,padilla2006}. As a consequence, to describe 
the structure of Kr isotopes, it is necessary to extend the IBM 
framework so as to include the effect of particle-hole excitations. To 
this end, we have adapted the configuration mixing technique  developed 
by Duval and Barrett \cite{duval1981,duval1982}. Within this context, 
shell-model-like $2k$-particle-$2k$-hole ($k=0,1,2,\ldots$) 
configurations are associated with boson spaces comprising $n_b+2k$ 
bosons. The different boson subspaces are allowed to mix via an 
interaction term that does not preserve the boson number. The 
configuration-mixing IBM Hamiltonian is then diagonalized in the space 
$[n_b]\oplus [n_b+2]\oplus [n_b+4]\oplus\cdots$, with $[n_b+2k]$ being 
the unperturbed boson subspace. As in several earlier calculations made
in the same mass region (e.g.,
\cite{duval1983,padilla2006,nomura2017ge}), we have  considered proton
particle-hole excitations across the $Z=28$ major shell gap. 
Moreover, as will be shown below, the Gogny-HFB energy surfaces display up to three 
mean-field minima. Those minima are sufficiently well defined so as to 
constrain the corresponding unperturbed IBM Hamiltonian and, therefore, 
we consider up to three configurations: the normal $0p-0h$ as well as 
the intruder $2p-2h$ and $4p-4h$ excitations.

The configuration mixing IBM Hamiltonian employed in this work reads
\begin{eqnarray} 
\label{eq:ham}
 \hat H = \hat H_{0} + (\hat H_{1} + \Delta_{1}) + (\hat H_{2} +
 \Delta_{2}) + \hat H_{01}^{\rm
 mix} + \hat H_{12}^{\rm mix}, 
\end{eqnarray}
where $\hat H_k$ ($k=0,1,2$) is the Hamiltonian for the unperturbed 
configuration space $[n_b+2k]$ while $\hat H^{\rm mix}_{kk+1}$ ($k=0,1$)
stands for the interaction mixing $[n_b+2k]$ and $[n_b+2(k+1)]$
spaces. In Eq.~(\ref{eq:ham}), $\Delta_1$ and $\Delta_2$ represent the
energy needed to excite one and two bosons from one major shell to the next.

For each configuration space, we have employed the simplest form of the 
IBM-1 Hamiltonian that still simulates the essential ingredients of the 
low-energy quadrupole dynamics, i.e., 
\begin{eqnarray}
\label{eq:ham-sg}
 \hat H_k = \epsilon_k\hat n_d+\kappa_k\hat Q\cdot\hat Q + \kappa^{\prime}_k\hat V_{ddd},
\end{eqnarray}
The first term in Eq.~(\ref{eq:ham-sg}) is the $d$-boson number
operator, with $\epsilon_k$ ($k=0,1,2$) being the single $d$-boson
energy in the $[n_b+2k]$ space. The second term represents 
the quadrupole-quadrupole interaction with strength 
parameter $\kappa_k$. The quadrupole operator $\hat Q$  reads $\hat
Q=s^{\dagger}\tilde d + d^{\dagger}s + \chi_k[d^{\dagger}\times\tilde d]^{(2)}$, where $\chi_k$ is
a parameter. On the other hand, the third term stands 
for the most relevant 
three-body interaction with strength  
$\kappa^{\prime}_k$.  This term is required to describe $\gamma$-soft 
systems \cite{vanisacker1981,nomura2012tri} and 
takes the form
\begin{eqnarray}
\label{eq:ddd}
 \hat V_{ddd} = [[d^{\dagger}\times d^{\dagger}\times d^{\dagger}]^{(3)}\times [[\tilde d\times\tilde d\times\tilde d]^{(3)}]^{(0)}.
\end{eqnarray}
The mixing interaction $\hat H_{kk+1}^{\rm mix}$ ($k=0$ or 1) reads
\begin{eqnarray}
 \hat H^{\rm mix}_{kk+1}=\omega^s_{k} s^{\dagger}s^{\dagger} + \omega^d_{k} d^{\dagger}\cdot d^{\dagger} + (h.c.), 
\end{eqnarray}
where $\omega^s_{k}$ and $\omega^d_k$ are strength parameters. For
simplicity, we have assumed  $\omega^s_k=\omega^d_k\equiv\omega_k$.
There is no direct coupling between the $[n_b]$ and $[n_b+4]$ spaces
with the two-body  interactions.

To associate the configuration-mixing IBM Hamiltonian of Eq.~(\ref{eq:ham})
with the corresponding Gogny-HFB energy surface, an extended 
boson coherent state has been introduced \cite{frank04}: 
\begin{eqnarray}
\label{eq:coherent}
|n_0,(\beta_0,\gamma_0)\rangle\oplus
|n_1,(\beta_1,\gamma_1)\rangle\oplus |n_2,(\beta_2,\gamma_2)\rangle, 
\end{eqnarray}
where $n_k=n_b+2k$ ($k=0,1,2$). For each unperturbed configuration 
space $|n_k,(\beta_k,\gamma_k)\rangle$ ($k=0,1,2$), the coherent state 
is taken in the form
\begin{eqnarray}
\label{eq:coherent-unp}
&&|n_k,(\beta_k,\gamma_k)\rangle=\frac{1}{\sqrt{n_{k}!}}\times \nonumber \\
&&(
(s^{\dagger}+\beta_k\cos{\gamma_k}d^{\dagger}_0+\frac{1}{\sqrt{2}}\beta_k\sin{\gamma_k
}(d^{\dagger}_{+2}+d^{\dagger}_{-2})
)
)^{n_k}|0\rangle
\end{eqnarray}
where $|0\rangle$ denotes the inert core. For each unperturbed 
configuration $[n_b+2k]$, the boson analogs of the quadrupole 
deformation parameters $\beta$ and $\gamma$ are denoted by $\beta_k$ 
and $\gamma_k$, respectively \cite{BM}. They are assumed to be in 
correspondence with the ones of the the Gogny-HFB by means of a linear 
dependence with $\beta_k=C_k\beta$ and $\gamma_k=\gamma$. The constants  
$C_k$ are  also determined  by fitting  the (fermionic) Gogny-HFB 
energy surface to the (bosonic) IBM one by requiring  that the position 
of the minimum is reproduced for each unperturbed configuration.

The expectation value of the total Hamiltonian $\hat H$ in the coherent 
state Eq.~(\ref{eq:coherent}) leads to a $3\times 3$ matrix \cite{frank04}: 
\begin{eqnarray}
\label{eq:pes}
  {\cal E}=\left(
\begin{array}{ccc}
E_{0}(\beta,\gamma) & \Omega_{01}(\beta) & 0 \\
\Omega_{01}(\beta) & E_{1}(\beta,\gamma)+\Delta_{1} & \Omega_{12}(\beta) \\
0 & \Omega_{12}(\beta) & E_{2}(\beta,\gamma)+\Delta_{2} \\
\end{array}
\right), 
\end{eqnarray}
with diagonal and off-diagonal elements accounting for the expectation 
values of the unperturbed and mixing terms, respectively. The three 
eigenvalues of ${\cal E}$ correspond to specific energy surfaces. It is 
customary to take the lowest-energy one \cite{frank04} as the IBM 
energy surface. Both $E_k(\beta,\gamma)$ and $\Omega_{kk+1}(\beta)$ are 
computed analytically. Their  expressions can be  found in 
Ref.~\cite{nomura2017ge}.


\subsection{Derivation of the IBM parameters: the fitting procedure\label{sec:mapping}}


The  Hamiltonian in Eq.~(\ref{eq:ham}) contains 16 parameters. They have 
been determined along the following lines:
 \begin{enumerate}
 \item[(i)] The unperturbed Hamiltonians are determined by using
	    the procedure of
	    Refs.~\cite{nomura2008,nomura2010,nomura2016zr}:
	    each diagonal matrix element $E_k(\beta,\gamma)$ in
	    Eq.~(\ref{eq:pes}) is fitted to reproduce the topology of
	    the Gogny-HFB energy surface in the neighborhood of the
	    corresponding minimum. 
	    The normal $[n_b]$ configuration is assigned to the HFB minimum
	    with the smallest $\beta$ deformation, the 
	    $[n_b+2]$ configuration is assigned to
	    the minimum with the second smallest $\beta$ deformation and
	    the $[n_b+4]$ configuration is associated to the 
	    minimum with the third
	    smallest $\beta$ deformation. 
	    In this way, each 
	    unperturbed Hamiltonian is determined
	    independently.	    	    		       	   	    
\item[(ii)] The energy offset $\Delta_{k+1}$ ($k=0,1$) is determined so
	    that the energy difference 
              between the two minima of the Gogny-HFB energy surface, associated 
	      with the $[n_b+2k]$ and
	    $[n_b+2(k+1)]$ configurations,  is
	          reproduced. 	    		  		  				  
\item[(iii)] The strength parameter $\omega_{kk+1}$ ($k=0,1$)
	     of the mixing interaction term $\hat H^{\rm mix}_{kk+1}$
	     is determined so as to reproduce the shapes of the barriers
	     between the minima  corresponding to the $[n_b+2k]$ and 
	    $[n_b+2(k+1)]$ configurations
	     \cite{nomura2012sc,nomura2013hg}. Steps (ii) and (iii) are
	     repeated until the best match is obtained between the HFB
	     and IBM energy surfaces. 	     
\end{enumerate}
We have assumed that the boson-number dependence of the
$\kappa$ parameter is consistent with  earlier IBM calculations 
\cite{OAI,nomura2008}, i.e., $\kappa$ decreases in magnitude as a
function of $n_b$, to determine  the parameters of the 
unperturbed Hamiltonians. In  step (i), the link of the unperturbed configurations with
the deformed minima is based on the assumption that the
interpretation of shape coexistence in the
neutron-deficient lead region 
\cite{bengtsson1987,bengtsson1989,nazarewicz1993} also holds here. 
In the case of mercury nuclei, for
instance, the $0^+_1$ ground state is associated with a weakly-deformed
oblate shape and the intruder $0^+_2$ state with a prolate shape with a
larger $\beta$ deformation \cite{bengtsson1987,nazarewicz1993}. 
Obviously, this
assumption can only be tested {\it{a posteriori}} as a function
of the results obtained for the considered nuclei.

Once the IBM parameters are determined for each Kr nucleus, the
Hamiltonian $\hat H$ is diagonalized in the 
$[n_b]\oplus [n_b+2]\oplus [n_b+4]$ space by using the code IBM-1
\cite{IBM1}. The IBM wave functions resulting from the diagonalization
are then used to compute electromagnetic properties, including E2 and E0
transitions, that could be considered signatures of shape coexistence
and shape transitions. The $B(E2)$ transition probability reads
\begin{eqnarray}
 B(E2; J_i\rightarrow J_f)=\frac{1}{2J_i+1}|\langle J_f||\hat T^{(E2)}||J_i\rangle|^2, 
\end{eqnarray}
where $J_i$ and $J_f$ are the spins of the initial and final states,
respectively.  On the other hand, the  $\rho^2({E0})$ values are computed
as 
\begin{eqnarray}
\label{eq:rhoe0}
 \rho^2(E0; 0^+_i\rightarrow 0^+_f) = \frac{Z^2}{R_0^4}|\langle 0^+_f||\hat T^{(E0)}||0^+_i\rangle|^2
\end{eqnarray}
where $R_0=1.2\,A^{1/3}$ fm. 

The E0 and E2 operators take the form 
$\hat T^{(E0)}=\sum_{n=0,1}(e_{0,n}^s\hat n_s + e_{0,n}^d\hat n_d)$ and 
$\hat T^{(E2)}=\sum_{n=0,1}e_{2,n}\hat Q$, respectively. 
For the  effective charges for the E0 operator we have assumed
$e_{0,0}^s=e_{0,1}^s=e_{0,2}^s\equiv e_{0}^s$ as well as 
$e_{0,0}^d=e_{0,1}^d=e_{0,2}^d\equiv e_{0}^d$. 
Also, the ratio $e_{0}^d/e_{0}^s=1.4$ has been assumed so as to obtain an overall
agreement with the  experimental trend of the
$\rho^2(E0; 0^+_2\rightarrow 0^+_1)$ values around $N=40$. 
The remaining parameter $e_0^s$ is fitted to reproduce the experimental
$\rho^2(E0; 0^+_2\rightarrow 0^+_1)$ value for  
$^{76}$Kr. For the E2 effective charges, we have assumed  the ratios 
$e_{2,1}/e_{2,0}=\kappa_{1}/\kappa_{0}$ and
$e_{2,2}/e_{2,0}=\kappa_{2}/\kappa_{0}$, based on the fact that both the 
effective charge and quadrupole interaction are proportional to the mean-square proton
radius \cite{duval1981,duval1982}. We have then fitted the overall
factor $e_{2,0}$ to the experimental $B(E2; 2^+_1\rightarrow 0^+_1)$
value  for $^{76}$Kr \cite{data}.


\section{Results\label{sec:results}}


In this section, we discuss the results of our calculations 
for the selected set of Kr isotopes. 
First, in Sec.~\ref{sec:pes}, we discuss 
the $(\beta,\gamma)$-deformation energy surfaces obtained from the SCMF
calculations as well as the mapped IBM ones. 
The IBM parameters derived via the mapping procedure 
and the configurations employed in the calculations are 
presented in Sec.~\ref{sec:config}. 
In Secs.~\ref{sec:energy}
and \ref{sec:transition}, we discuss spectroscopic properties such as
the systematics of the energy spectra and the transition rates predicted
in our calculations in comparison with the available
experimental data. The detailed spectroscopy of a selected sample of Kr
isotopes is discussed in Sec.~\ref{sec:detail}.
Finally, in Sec.~\ref{sec:uncertainty}, we consider the sensitivity 
of the results with respect to the underlying EDF used in the mapping procedure.


\subsection{Deformation energy surfaces\label{sec:pes}}



\begin{figure*}[htb!]
\begin{center}
\includegraphics[width=\linewidth]{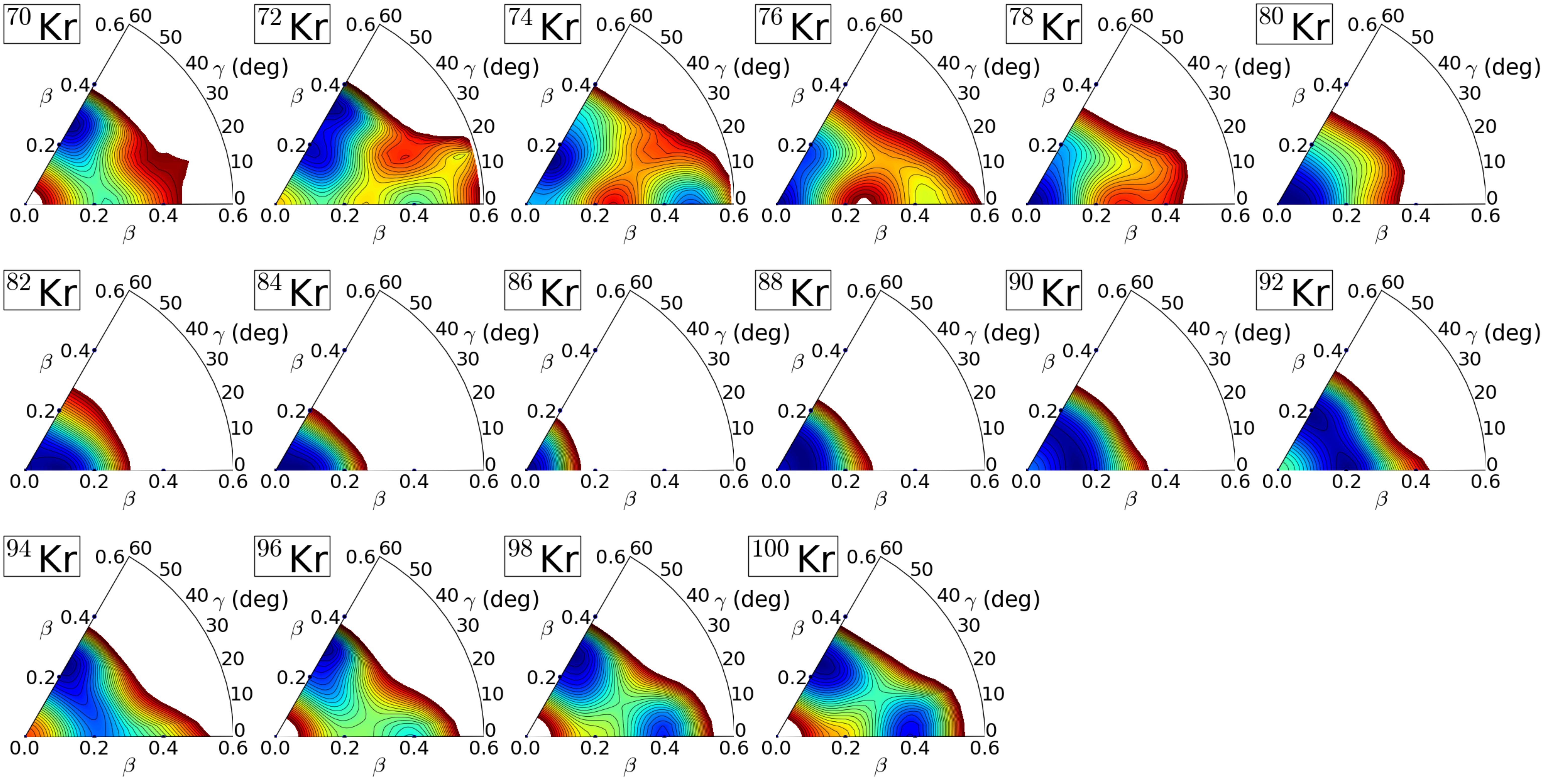}
\caption{(Color online) SCMF $(\beta,\gamma)$-deformation energy
 surfaces for the $^{70-100}$Kr nuclei, obtained with the Gogny-D1M EDF. The energy difference 
between neighboring contours is 100 keV. }
\label{fig:pes-hfb}
\end{center}
\end{figure*}


\begin{figure}[htb!]
\begin{center}
\includegraphics[width=0.7\columnwidth]{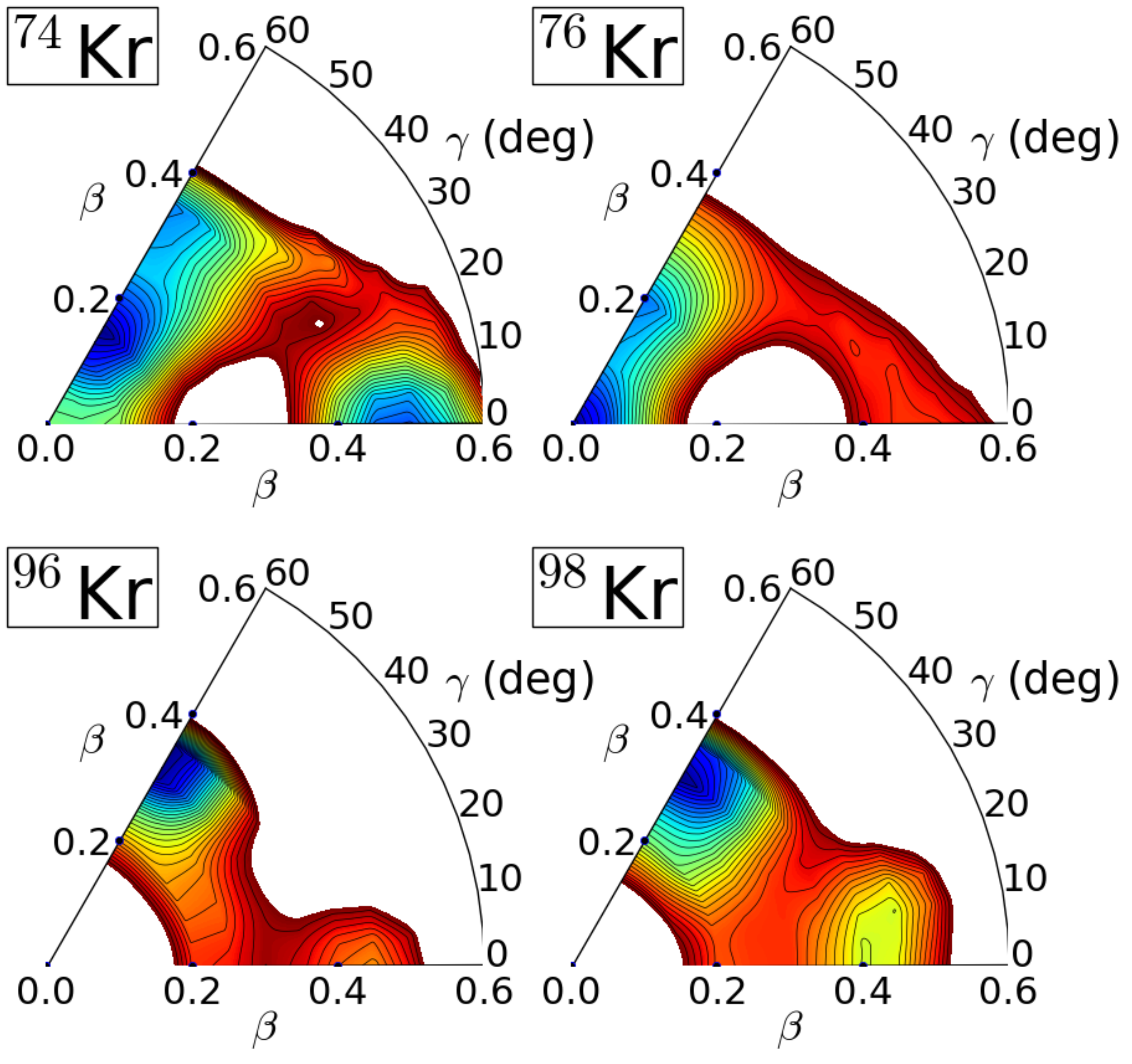}
\caption{(Color online) The same as in Fig.~\ref{fig:pes-hfb}, but for 
 $^{74}$Kr, $^{76}$Kr, $^{96}$Kr and $^{98}$Kr, computed with the DD-PC1 EDF.}
\label{fig:pes-hfb-ddpc1}
\end{center}
\end{figure}


\begin{figure*}[htb!]
\begin{center}
\includegraphics[width=\linewidth]{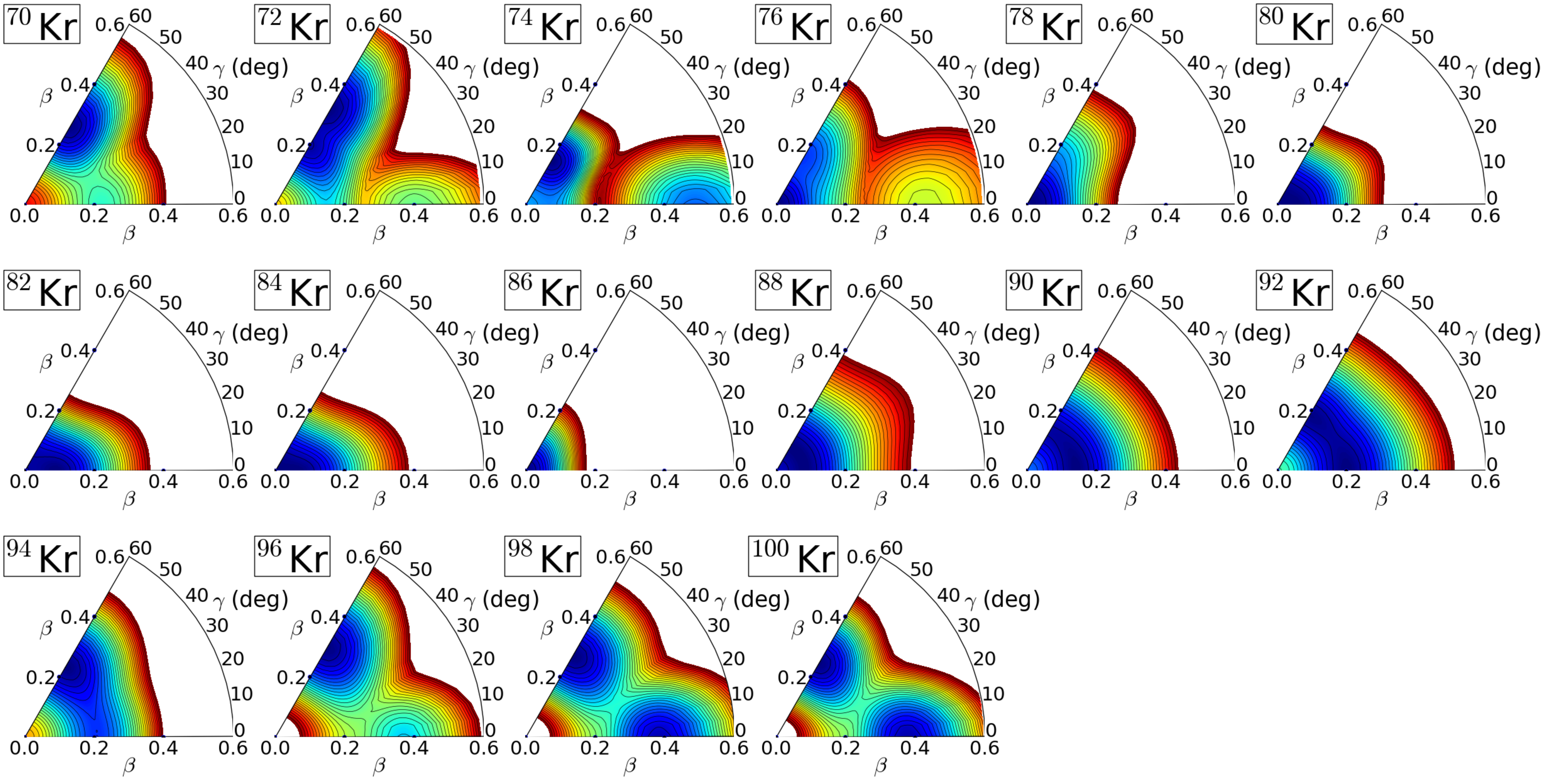}
\caption{(Color online) The same as in Fig.~\ref{fig:pes-hfb}, 
but for the mapped IBM energy surfaces.}
\label{fig:pes-mapped}
\end{center}
\end{figure*}

The Gogny-D1M HFB energy surfaces
are depicted  in Fig.~\ref{fig:pes-hfb}
for the studied $^{70-100}$Kr isotopes. In the case 
of $^{70}$Kr, one observes an absolute oblate and 
a secondary prolate minima. The energy surface 
obtained for $^{72}$Kr exhibits a complex topology
with two oblate ($\beta\approx 0.2$ and $\approx 0.3$)
and a prolate ($\beta\approx 0.4$) minima.
For $^{74}$Kr, the two oblate minima ($\beta=0.04$ and 0.15) are much
softer in $\beta$, while the prolate one becomes more pronounced. 
Thus, $^{74}$Kr presents one of the best examples for the
prolate-oblate shape coexistence in this region of the nuclear chart. 
In the case of $^{76}$Kr, we find a spherical global minimum 
that could be associated with the 
neutron $N=40$ sub-shell closure
and a very 
shallow
oblate minimum around $\beta=0.17$. On the other hand, the
prolate minimum  becomes less pronounced. For the nucleus 
$^{78}$Kr, we have obtained an almost 
invisible prolate minimum while for $^{80-86}$Kr only 
a single nearly-spherical
minimum is found reflecting, the effect of the 
$N=50$  shell closure. On the neutron-rich side with $N>50$, one  realizes that the
Gogny-D1M surfaces for $^{88-92}$Kr exhibit a 
pronounced $\gamma$ softness, almost flat independent of $\gamma$
deformation. A $\gamma$-soft oblate minimum 
develops for $^{94}$Kr. In the case of $^{96-100}$Kr, the prolate 
local minimum appears at 
$\beta\approx 0.4$. Those  nuclei exhibit a spectacular prolate-oblate shape
coexistence, similarly to  $^{72-76}$Kr on the neutron-deficient side.

Let us mention, that similar energy surfaces have
been obtained for the considered nuclei
with the Gogny-D1S and D1N EDFs. However, we observe certain quantitative 
differences between the Gogny-D1M and the relativistic
EDFs, especially for those nuclei around $N=40$ and $N=60$. Hence we present 
in Fig.~\ref{fig:pes-hfb-ddpc1} the SCMF energy surfaces obtained for
$^{74,76}$Kr and $^{96,98}$Kr using  
relativistic Hartree-Bogoliubov calculations  based on the DD-PC1 EDF. 
The DD-PC1 EDF provides stiffer energy surfaces with much higher
barriers between the minima than the Gogny-D1M ones. 
We have also confirmed that there are no significant differences between the
energy surfaces obtained with the relativistic DD-PC1 and DD-ME2 mean-field
Lagrangians.

Finally we present in Fig.~\ref{fig:pes-mapped} the mapped IBM energy
surfaces based on the Gogny-D1M  ones already  shown in Fig.~\ref{fig:pes-hfb}.
The comparison between the Gogny-D1M and IBM surfaces reveals that the latter mimic 
key features of the former in 
the neighborhood of the minima (their locations and depths, the curvatures along the $\beta$ 
and $\gamma$ directions).  As in previous works \cite{nomura2016zr,nomura2017ge}, the 
IBM surfaces look simpler than the mean-field ones. For instance, in the 
region far from each minimum the IBM surfaces
become too flat. Such a discrepancy can be attributed to the 
simplified form of the considered IBM Hamiltonian and/or to the limited boson model space 
built only on the valence nucleons. Nevertheless, as will be shown later on in 
this paper, the low-lying collective states are determined mainly by
the configurations around the minima, while the regions far from the
minima are dominated by single-particle degrees of freedom. 
This is the reason why we have tried to reproduce the topology of the
Gogny-D1M energy  surfaces only in the neighborhood of the corresponding minima.


\subsection{Configurations and derived parameters\label{sec:config}}


The ($\beta,\gamma$)-coordinates  on
the Gogny-D1M energy surfaces 
associated with the unperturbed IBM Hamiltonians of the $[n_b]$, $[n_b+2]$
and $[n_b+4]$ configurations are given in Table~\ref{tab:config}. Let us mention, that
the assignment of the unperturbed
configurations for  $^{70}$Kr and $^{94}$Kr does not follow
the rule mentioned in Sec.~\ref{sec:mapping} [step (i)]. For those 
nuclei, the normal
$[n_b]$ configuration is assigned to the oblate minimum while 
the $[n_b+2]$ configuration is assigned to the prolate
one with smaller $\beta$ value than the former. 
The reason is that
we assume that the 
intrinsic
structure of each unperturbed Hamiltonian does not change too much from
one nucleus to the next. As can be seen from the table, the 
assignment of the $[n_b]$ and $[n_b+2]$ configurations 
to oblate and prolate shapes in the nuclei 
$^{72}$Kr and $^{92,96}$Kr is similar to the cases 
of $^{70}$Kr
and $^{94}$Kr, respectively. 
It is also apparent from the table that the 
number of configurations included in the model space differs from
nucleus to nucleus. Let us stress, that intruder configurations are 
included in our calculations depending on whether the curvatures around
the HFB minimum in both $\beta$ and $\gamma$ directions 
are large enough to uniquely determine the corresponding unperturbed
Hamiltonian.

\begin{table}[htb]
\caption{\label{tab:config}
The ($\beta,\gamma$) coordinates on 
the Gogny-D1M energy surfaces associated with the unperturbed
IBM Hamiltonians in the $[n_b]$, $[n_b+2]$ and $[n_b+4]$ configurations.}
\begin{center}
\begin{tabular}{lcccccc}
\hline\hline
          & $[n_b]$ & $[n_b+2]$ & $[n_b+4]$ \\
\hline
$^{70}$Kr & (0.26, $60^{\circ}$) & (0.23, $0^{\circ}$) & - \\
$^{72}$Kr & (0.19, $60^{\circ}$) & (0.32, $60^{\circ}$) & (0.40,
	     $0^{\circ}$) \\
$^{74}$Kr & (0.04, $60^{\circ}$) & (0.15, $60^{\circ}$) & (0.48,
	     $0^{\circ}$) \\
$^{76}$Kr & (0.0, $0^{\circ}$) & (0.17, $60^{\circ}$) & (0.43,
	     $0^{\circ}$) \\
$^{78}$Kr & (0.0, $0^{\circ}$) & (0.15, $60^{\circ}$) & - \\
$^{80}$Kr & (0.04, $0^{\circ}$) & - & - \\
$^{82}$Kr & (0.11, $0^{\circ}$) & - & - \\
$^{84}$Kr & (0.06, $0^{\circ}$) & - & - \\
$^{86}$Kr & (0.0, $0^{\circ}$) & - & - \\
$^{88}$Kr & (0.08, $0^{\circ}$) & - & - \\
$^{90}$Kr & (0.14, $0^{\circ}$) & - & - \\
$^{92}$Kr & (0.19, $60^{\circ}$) & (0.21, $0^{\circ}$) & - \\
$^{94}$Kr & (0.25, $60^{\circ}$) & (0.21, $0^{\circ}$) & - \\
$^{96}$Kr & (0.31, $60^{\circ}$) & (0.40, $0^{\circ}$) & - \\
$^{98}$Kr & (0.28, $60^{\circ}$) & (0.40, $0^{\circ}$) & - \\
$^{100}$Kr & (0.25, $60^{\circ}$) & (0.38, $0^{\circ}$) & - \\
\hline\hline
\end{tabular}
\end{center}
\end{table}


\begin{figure}[htb!]
\begin{center}
\includegraphics[width=\columnwidth]{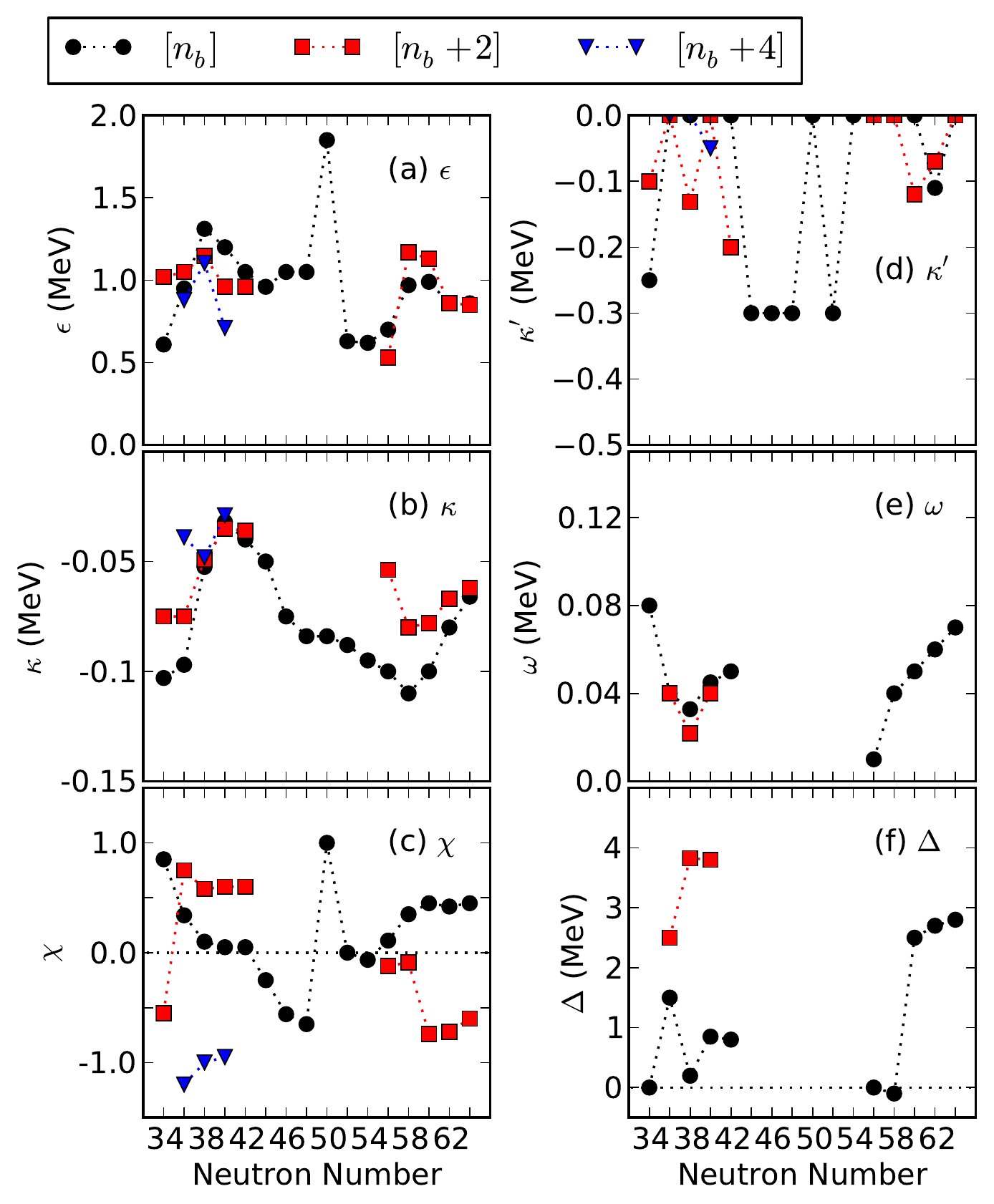}
\caption{(Color online) Derived IBM parameters for the
 $[n_b]$, $[n_b+2]$ and $[n_b+4]$ configurations as functions of 
the neutron number.}
\label{fig:para}
\end{center}
\end{figure}

The parameters of the IBM Hamiltonian Eq.~(\ref{eq:ham}) are plotted 
in Fig.~\ref{fig:para} as functions of the neutron number $N$. 
Since, as already mentioned, the configuration space is different from nucleus
to nucleus, in some Kr nuclei not all the parameters appear in the figure. 
For instance, values of the parameters  $\omega$ [panel (e)] and $\Delta$ [panel (f)] are not
plotted for those nuclei with $N=44-56$ as the configuration mixing was 
not performed for them. 

In the case of the unperturbed Hamiltonians
[panels (a) to (d)] those parameters reflect the structural evolution
along the considered isotopic chain. For example, the $\epsilon$ value for the
$[n_b]$ configuration becomes larger towards the neutron sub-shell
closure $N\approx 40$ [see, panel
(a)] though this is not taken into account explicitly in the
model space we have employed in this study. On the other hand, the parameter
$\kappa$ [panel (b)] is much larger than the one employed in the 
IBM-1 phenomenology \cite{bai2016}. Such a large 
$\kappa$ value is required to reproduce 
the curvature around the minimum of the Gogny-D1M energy
surface. The positive (negative) values of the parameter $\chi$ [panel (c)]
correspond to  oblate (prolate) shapes. 

The $\hat V_{ddd}$ term in Eq.~(\ref{eq:ddd}) is relevant for $\gamma$
softness, as it gives rise to a triaxial minimum with $\gamma\neq
0^{\circ}$ and/or 60$^{\circ}$ \cite{vanisacker1981}. 
On the other hand, the Gogny-D1M energy surfaces in 
Fig.~\ref{fig:pes-hfb} suggest that none of the considered Kr nuclei exhibits
a triaxial mean-field
minimum. 
Therefore, we have assumed that the effect of this term is
rather perturbative in this particular study, and the strength parameter
$\kappa^{\prime}$ has 
been introduced only for those configurations
corresponding to minima that are relatively soft along the $\gamma$
direction so that mainly the dependence of the energy 
surfaces on $\gamma$  is reproduced. 
We have also verified that the inclusion of the three-body boson term
$\hat V_{ddd}$ improves only little the description of the energy spectra. 

The  behavior of the 
mixing strength $\omega$ [panel (e)]
and 
the energy offset $\Delta$ [panel (f)] around 
$N=36$ and 60 correspond to the significant change expected 
in the nuclear structure around those neutron numbers.

\subsection{Systematics of excitation spectra\label{sec:energy}}


\begin{figure}[htb!]
\begin{center}
\includegraphics[width=\columnwidth]{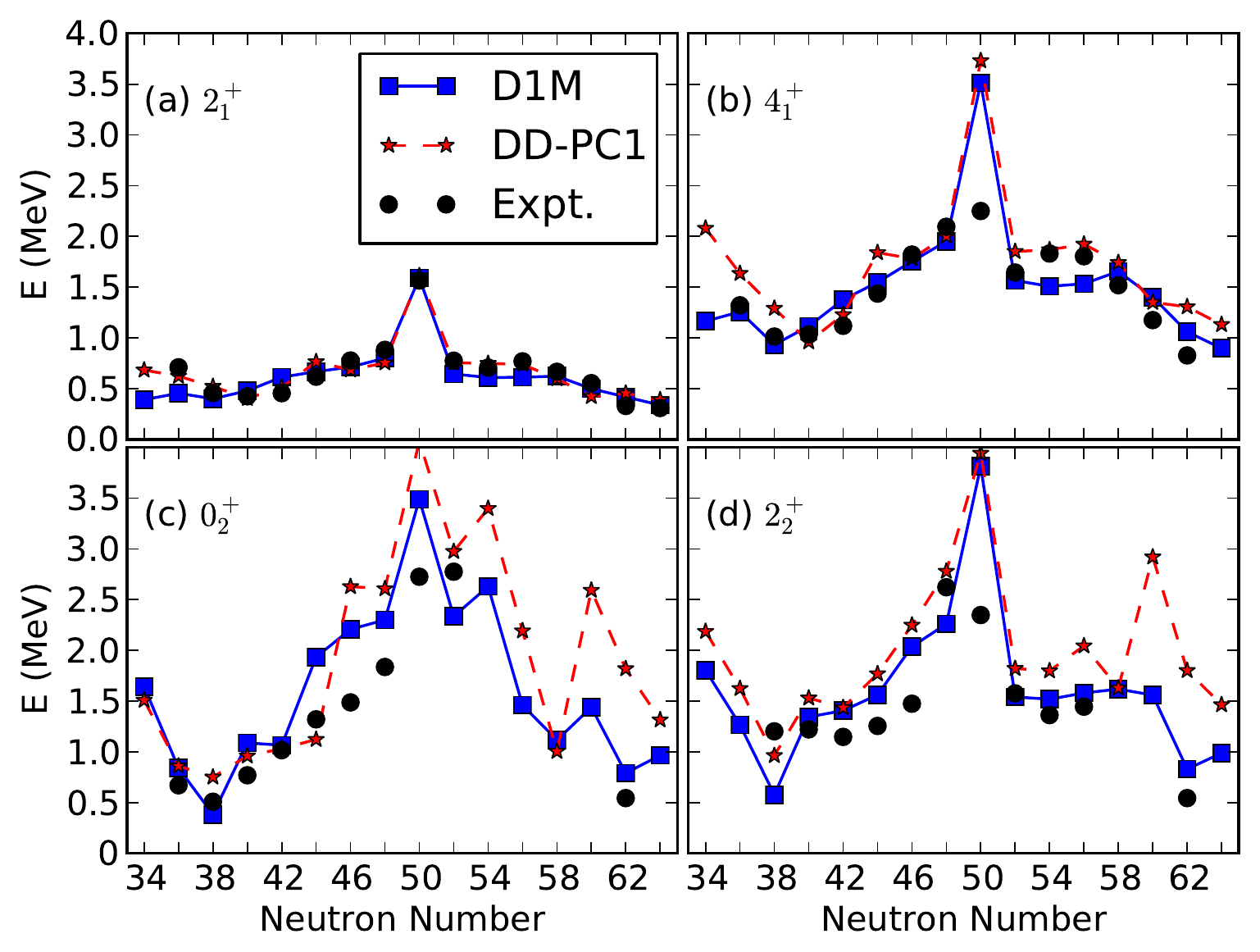}
\caption{(Color online) Experimental
 \cite{data,albers2013,rzkacaurban2017,dudouet2017,flavigny2017} and computed excitation
 spectra for 
 the $2^+_1$, $4^+_1$, $0^+_2$ and $2^+_2$ states in  $^{70-100}$Kr
 as functions of $N$. The theoretical results are obtained with the Gogny-D1M and
 relativistic DD-PC1 EDFs.}
\label{fig:energy}
\end{center}
\end{figure}

Even though the analysis of the $(\beta,\gamma)$-deformation energy
surfaces provides useful insights into both the shape transition
and shape coexistence phenomena in the studied Kr isotopes, a more 
quantitative analysis should go beyond 
the mean-field level to examine  spectroscopic 
properties such as the  excitation spectra and transition rates which 
can be directly compared with the available 
experimental data. In this and the following Sec.~\ref{sec:transition}, we turn our 
attention to those properties.

The excitation energies of the 
$2^+_1$ [panel (a)], $4^+_1$ [panel (b)], $0^+_2$ [panel (c)] and
$2^+_2$ [panel (d)] states
are plotted in Fig.~\ref{fig:energy} as functions of the neutron number $N$.
The results obtained with the parametrization D1M of the Gogny-EDF
are compared with those obtained using the 
DD-PC1 relativistic mean-field Lagrangian as well as with the available
experimental data \cite{data,albers2013,rzkacaurban2017,dudouet2017,flavigny2017}. The  
fraction of the three
configurations $[n_b]$, $[n_b+2]$ and $[n_b+4]$ 
in the wave functions of the $0^+_1$, $2^+_1$, $0^+_2$
and $2^+_2$ states are given in Table~\ref{tab:frac}.
 
The energy spectra, computed with the D1M and DD-PC1 EDFs display 
a reasonable agreement with the experimental data. The 
$E(2^+_1)$
excitation energy can be regarded as 
one of the basic quantities  signalling  a shape/phase transition
\cite{CasBook,cejnar2010}. The predicted $E(2^+_1)$  energies, shown in 
panel (a) of Fig.~\ref{fig:energy}, nicely follow the 
experimental trend though at $N=36$ our calculations 
rather underestimate the experiment. Note, that the 
lowering of the $E(2^+_1)$ towards the midshells, on both 
the neutron-deficient  ($N\approx 40$) and the 
neutron-rich ($N\approx 64$) sides, signals the 
emergence of quadrupole collectivity. Furthermore, the 
decrease of the predicted $E(2^+_1)$ energies on the 
neutron-rich side agrees well with 
the smooth onset of deformation suggested by  recent
experiments \cite{albers2012,albers2013}.  Similar 
results are obtained for the 
$E(4^+_1)$ excitation energies [panel (b)]. However, they
overestimate the experimental data at $N=50$ due to 
the limited IBM space consisting only of $s$ and $d$ bosons.
Those results also indicate the need of including 
$J^{\pi}=4^+$ (or $g$) bosons in our calculations. Work 
along these lines is in progress and will be reported
elsewhere.

The  $E(0^+_2)$ excitation energies are plotted in 
panel (c) of Fig.~\ref{fig:energy}. As can be seen, our 
calculations describe fairly well the  experimental data
around $N=40$ where a pronounced coexistence
between oblate and prolate shapes is suggested 
by the corresponding Gogny-D1M energy 
surfaces (see, Fig.~\ref{fig:pes-hfb}).
The predicted values overestimate
the experimental  ones from $N=44$ to 50 since configuration mixing
has not been  performed for those nuclei. Beyond the neutron shell closure $N=50$
one observes a lowering in the predicted 
energies towards $N=64$.

The  $E(0^+_2)$ values obtained with both the D1M and DD-PC1 EDFs display
 a peak at
$N=60$. This could be a consequence of the prolate local 
minimum  that emerges for $^{96}$Kr 
at $\beta\approx 0.4$ (see,
Fig.~\ref{fig:pes-hfb}). The $0^+_2$ state in this nucleus is mainly
made of the prolate configuration (see, Tables~\ref{tab:config} and \ref{tab:frac}). 
The Gogny-D1M result exhibits an abrupt decrease  from
$N=60$ to 62, where the prolate minimum becomes much more pronounced. 
A similar observation can be made for the systematics of the $E(2^+_2)$
excitation energies in panel (d). In the case of the 
DD-PC1 EDF, higher $E(0^+_2)$ and $E(2^+_2)$ excitation energies 
than those obtained with the Gogny-D1M EDF
are predicted for the neutron-rich Kr isotopes.
This difference can be mainly attributed to the different topology of
the corresponding energy surfaces. 
In fact, as we have already shown in Figs.~\ref{fig:pes-hfb} and
\ref{fig:pes-hfb-ddpc1}, the DD-PC1 surfaces are generally stiffer  than
the D1M ones.

\begin{table}[htb]
\caption{\label{tab:frac} Fraction (in units of percent) of the three configurations 
 $[n_0]$, $[n_1]$ and $[n_2]$ ($n_k=n_b+2k$) in the $0^+_1$,
 $0^+_2$, $2^+_1$ and $2^+_2$ wave functions of those Kr nuclei where
 configuration mixing has been performed in the present calculation. 
}
\begin{center}
\begin{tabular}{lcccccccccccc}
\hline\hline
\multirow{2}{*}{} 
& \multicolumn{3}{c}{$0^+_1$} & \multicolumn{3}{c}{$0^+_2$} &
 \multicolumn{3}{c}{$2^+_1$} & \multicolumn{3}{c}{$2^+_2$} \\
\cline{2-4}
\cline{5-7}
\cline{8-10}
\cline{11-13}
 & $[n_0]$ & $[n_1]$ & $[n_2]$ & $[n_0]$ & $[n_1]$ & $[n_2]$ & $[n_0]$ &
 $[n_1]$ & $[n_2]$ & $[n_0]$ & $[n_1]$ & $[n_2]$ \\
\hline
$^{70}$Kr & 89 & 11 & - & 18 & 82 & - & 94 & 6 & - & 23 & 77 & - \\
$^{72}$Kr & 58 & 42 & 0 & 42 & 58 & 0 & 43 & 56 & 0 & 57 & 43 & 0 \\
$^{74}$Kr & 5 & 77 & 18 & 2 & 17 & 82 & 2 & 54 & 44 & 2 & 44 & 54 \\
$^{76}$Kr & 23 & 74 & 3 & 68 & 28 & 3 & 12 & 85 & 4 & 13 & 68 & 19 \\
$^{78}$Kr & 56 & 44 & - & 44 & 56 & - & 42 & 58 & - & 49 & 51 & - \\
$^{94}$Kr & 83 & 17 & - & 17 & 83 & - & 87 & 13 & - & 20 & 80 & - \\
$^{96}$Kr & 93 & 7 & - & 10 & 90 & - & 96 & 4 & - & 19 & 81 & - \\
$^{98}$Kr & 60 & 40 & - & 42 & 58 & - & 50 & 50 & - & 56 & 44 & - \\
$^{100}$Kr & 32 & 68 & - & 68 & 32 & - & 21 & 79 & - & 79 & 21 & - \\
\hline\hline
\end{tabular}
\end{center}
\end{table}

\subsection{Systematics of E2 and E0 transition rates\label{sec:transition}}


\begin{figure}[htb!]
\begin{center}
\includegraphics[width=\columnwidth]{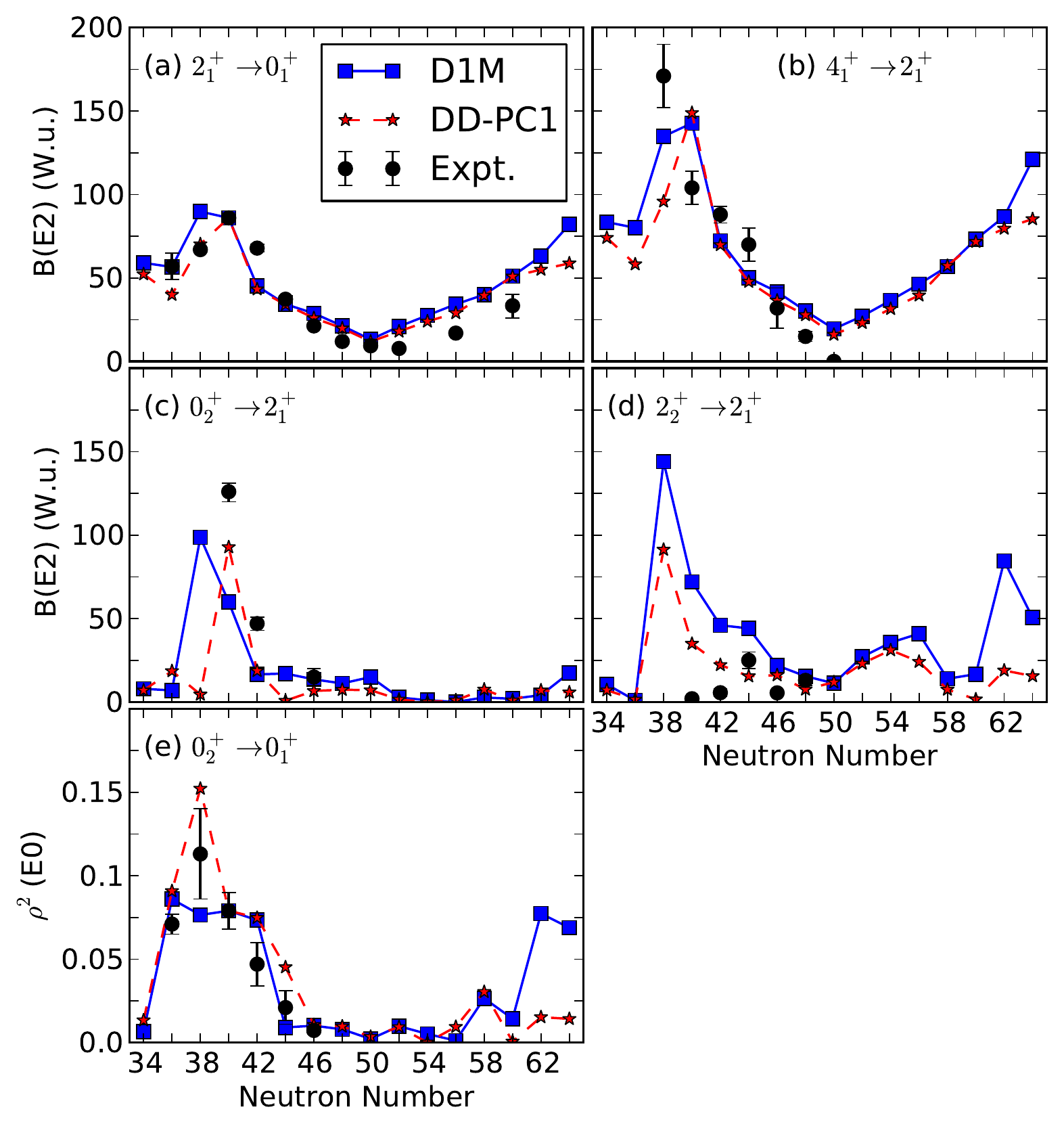}
\caption{(Color online) The experimental \cite{clement2007,data,kibedi2005} and
 theoretical $B(E2; 2^+_1\rightarrow 0^+_1)$ (a), 
 $B(E2; 4^+_1\rightarrow 2^+_1)$ (b), $B(E2; 0^+_2\rightarrow 2^+_1)$
 (c), and $B(E2; 2^+_2\rightarrow 2^+_1)$ (d) transition strengths (in
 Weisskopf units), and $\rho^2(E0; 0^+_2\rightarrow 0^+_1)$ values for
 the $^{70-100}$Kr nuclei depicted as functions of the neutron
 number. The theoretical calculations have been  performed based on the
 Gogny-D1M and relativistic  DD-PC1 EDFs.}
\label{fig:e2e0}
\end{center}
\end{figure}

In this section we discuss the 
systematics of the
$B(E2)$ and $\rho^2(E0)$ transition strengths. In Fig.~\ref{fig:e2e0} 
we have plotted, the 
experimental \cite{clement2007,data,kibedi2005} and
theoretical $B(E2; 2^+_1\rightarrow 0^+_1)$ [panel (a)], $B(E2; 4^+_1\rightarrow
2^+_1)$ [panel (b)], $B(E2; 0^+_2\rightarrow 2^+_1)$  [panel (c)], 
and $B(E2; 2^+_2\rightarrow 2^+_1)$ [panel (d)] transition strengths as well as
the $\rho^2(E0; 0^+_2\rightarrow 0^+_1)$ values [panel (e)], as functions of the neutron
number $N$. Results have been obtained with the Gogny-D1M and 
DD-PC1 EDFs.

The  $B(E2; 2^+_1\rightarrow 0^+_1)$ and $B(E2;4^+_1\rightarrow 2^+_1)$ transition 
probabilities agree reasonably well with the experimental data. They display the
well-known systematics signaling the development of collectivity, i.e.,  they
increase  when departing from the shell closure and become maximal around midshell.
On the other hand, the $B(E2; 0^+_2\rightarrow 2^+_1)$ transition
probabilities can be regarded as a measure of shape mixing. As can be seen from 
panel (c), the $B(E2; 0^+_2\rightarrow 2^+_1)$ values, obtained with both 
the  D1M and DD-PC1 EDFs, exhibit a peak around $N=40$ where the corresponding
mean-field energy surfaces display coexisting minima and their 
mixing is expected to be
strong. However, the theoretical $B(E2; 0^+_2\rightarrow 2^+_1)$ values for
$^{74,76}$Kr considerably underestimate the experimental
ones \cite{clement2007}.  Note, that the experimental value
$B(E2; 0^+_2\rightarrow 2^+_1)$ = 255$\pm 27$ W.u \cite{clement2007}  for
$^{74}$Kr  is too
large compared to the one obtained in our calculations and, therefore,
not shown in the figure. 
This is due to the fact (see, Table~\ref{tab:frac}) that  the
compositions of the $0^+_2$ and $2^+_1$
IBM wave functions are rather different. Furthermore, the small
$B(E2; 0^+_2\rightarrow 2^+_1)$
values obtained for neutron-rich Kr isotopes indicate 
that
there is almost no mixing between the $0^+_2$ and $2^+_1$ states.
A pronounced  difference between the D1M and DD-PC1 EDFs is observed 
in the case of $^{74}$Kr
for which
 the $B(E2; 0^+_2\rightarrow 2^+_1)$ value obtained with  the latter
is almost zero. As can be seen from Figs.~\ref{fig:pes-hfb} and 
\ref{fig:pes-hfb-ddpc1}, the
DD-PC1 energy surface for $^{74}$Kr  displays three minima within 1 MeV 
a structure, more complex than the corresponding Gogny-D1M one. Therefore, the
IBM Hamiltonian
used  in this study seems to be too simple to account for the large 
experimental
$B(E2; 0^+_2\rightarrow 2^+_1)$ value.

Experimental data are also  available for the 
$B(E2;2^+_2\rightarrow 2^+_1)$ transition probability.
They are depicted in panel (d) of Fig.~\ref{fig:e2e0}.
Our calculations, with both the Gogny-D1M 
and DD-PC1 EDFs, follow the
experimental trend from $N=44$ to 50. However, they 
overestimate the experimental
values for  $^{76,78}$Kr. As can be seen from Table~\ref{tab:frac}, the $2^+_1$ and $2^+_2$ wave
functions for those nuclei have a similar structure, leading to 
large E2 matrix elements.  

Finally, another signature of shape coexistence is provided 
by the $\rho^2(E0; 0^+_2\rightarrow 0^+_1)$ values \cite{heyde2011}. They
are compared in panel (e) with the experiment \cite{kibedi2005}.
Both the experimental and theoretical $\rho^2(E0; 0^+_2\rightarrow
0^+_1)$ values are notably large around $N=38-40$ 
signalling the shape coexistence in those isotopes. Note that, regardless
of the considered EDF, the agreement with the experimental data is rather
good.

\subsection{Detailed spectroscopy of selected isotopes\label{sec:detail}}


\begin{figure}[htb!]
\begin{center}
\includegraphics[width=\columnwidth]{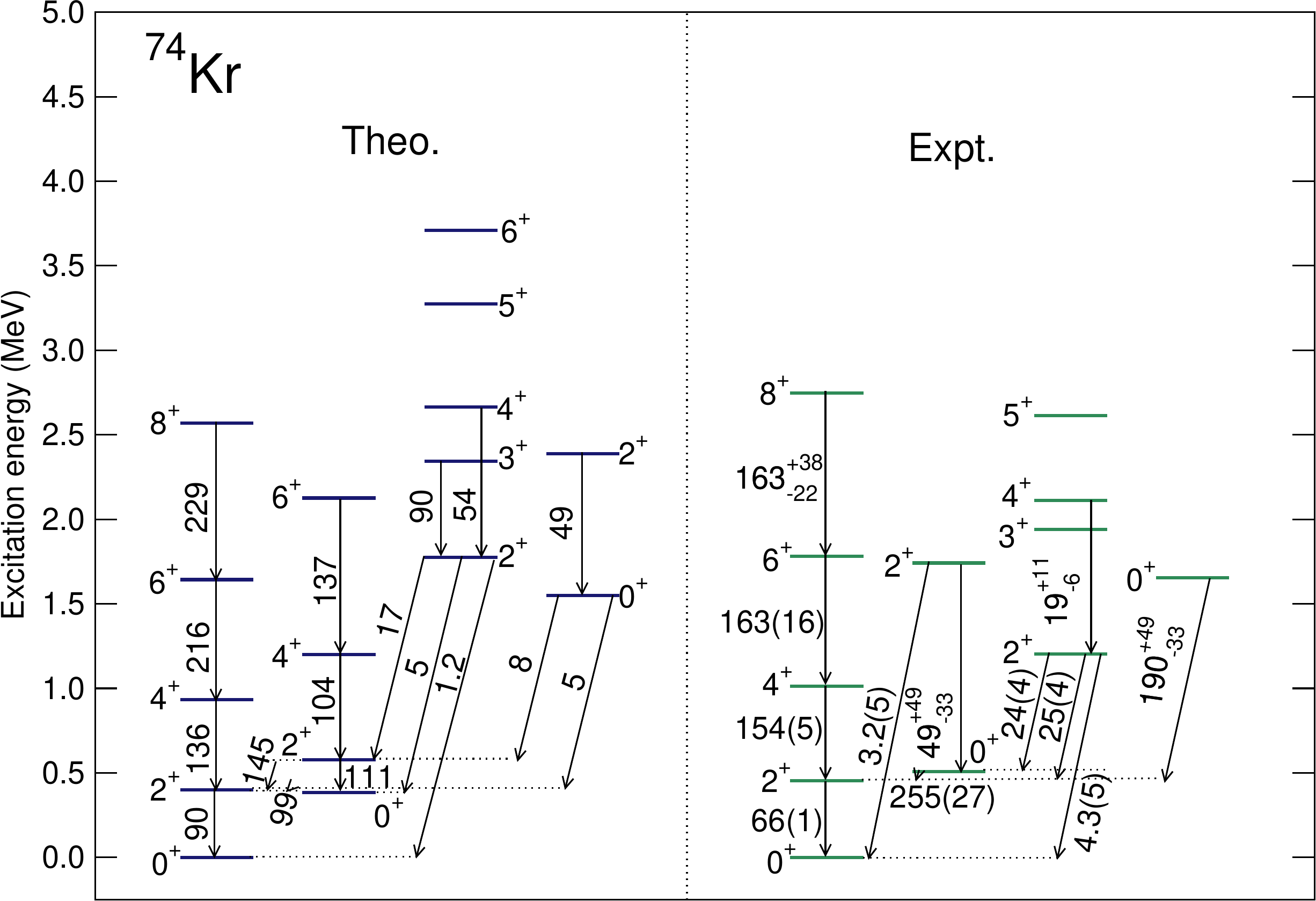} \\
\includegraphics[width=\columnwidth]{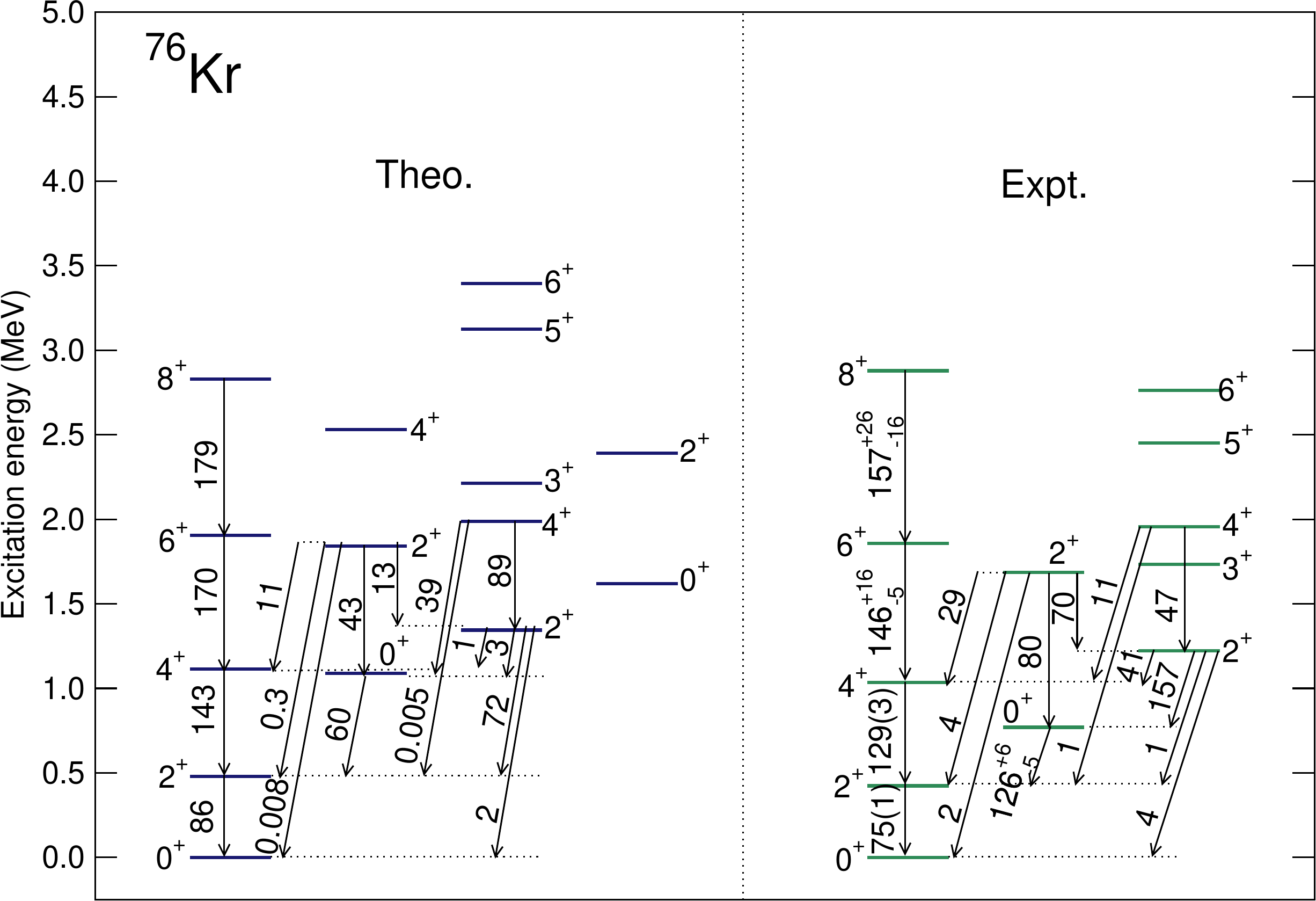}
\caption{(Color online) The theoretical low-energy excitation spectra
 and $B(E2)$ transition strengths (in W.u., indicated along arrows) of
 the $^{74}$Kr and  $^{76}$Kr isotopes obtained from the Gogny-D1M EDF,
 in comparison to  the available experimental data \cite{clement2007,data}.}
\label{fig:kr7476}
\end{center}
\end{figure}


\begin{figure}[htb!]
\begin{center}
\includegraphics[width=\columnwidth]{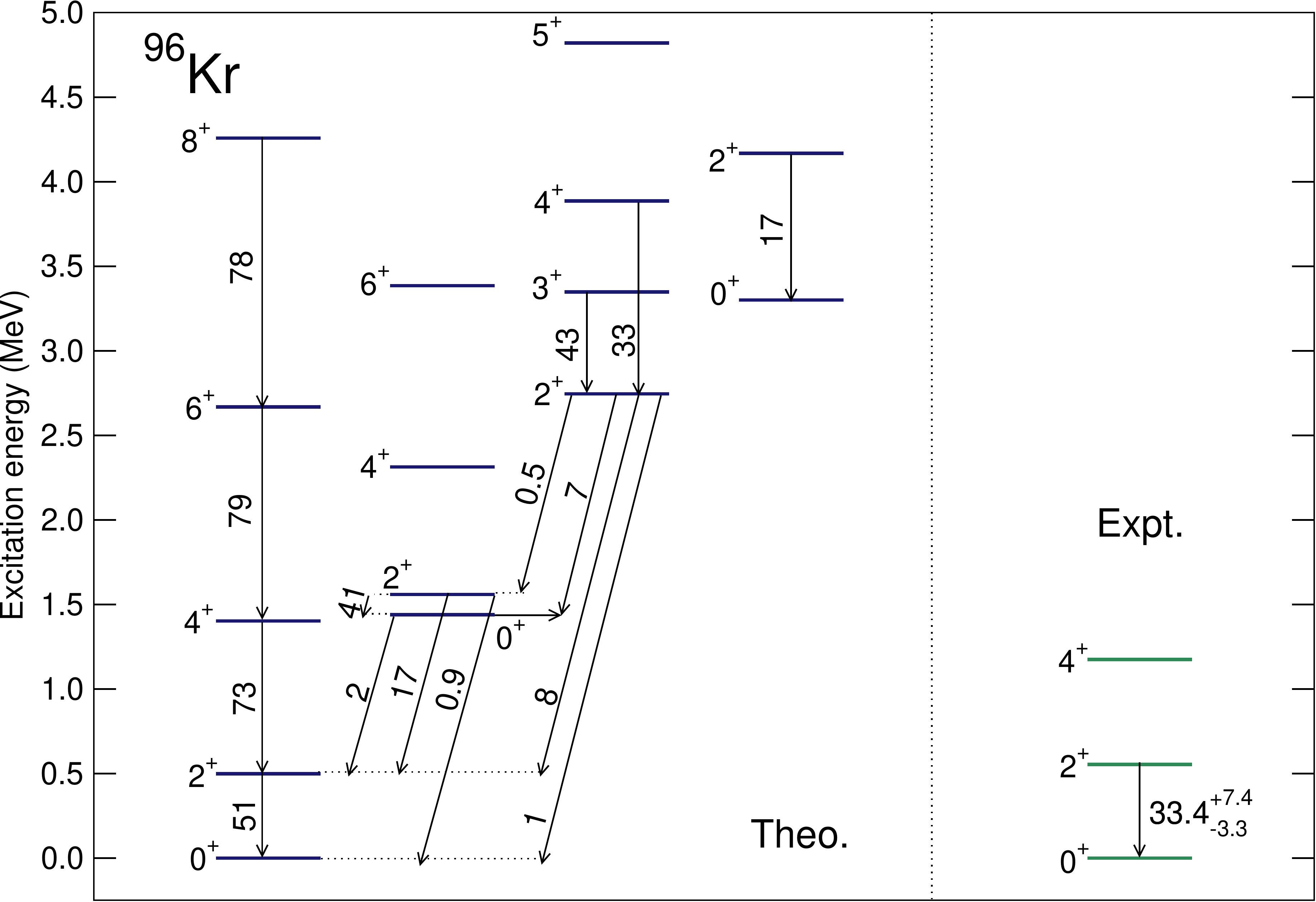} \\
\includegraphics[width=\columnwidth]{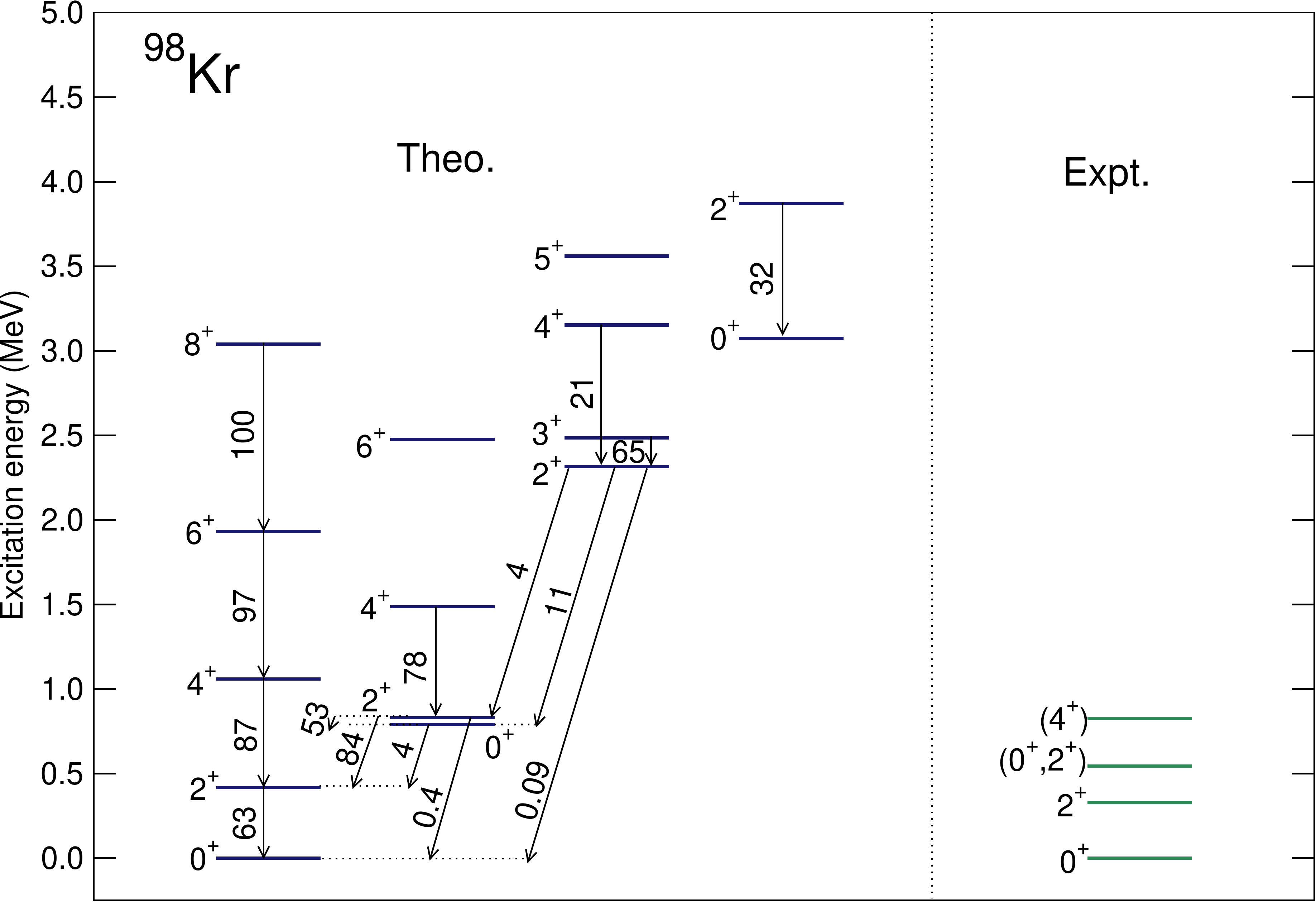}
\caption{(Color online) The same as in Fig.~\ref{fig:kr7476}, but for 
 $^{96}$Kr and $^{98}$Kr. The experimental data have been taken
 from Refs.~\cite{albers2013,dudouet2017,flavigny2017}.}
\label{fig:kr9698}
\end{center}
\end{figure}

We now turn our attention to a more detailed analysis of the low-energy
spectroscopy of individual nuclei. To this end, we consider the neutron-deficient
$^{74,76}$Kr and  the neutron-rich $^{96,98}$Kr  isotopes that exhibit 
a pronounced shape coexistence. The corresponding IBM states have been 
grouped into bands according to the dominant E2 decay patterns.

The Gogny-D1M  energy surfaces  for  $^{74,76}$Kr 
display coexisting spherical, oblate and prolate minima
(see, Fig.~\ref{fig:pes-hfb}). One of the most remarkable 
features of the spectra obtained for those nuclei is, the 
presence of  low-lying $0^+_2$ states
(see, Fig.~\ref{fig:energy}).  As can be seen from 
Fig.~\ref{fig:kr7476}, the low-energy excitation spectra
obtained for  $^{74,76}$Kr, with the Gogny-D1M 
EDF, agree reasonably well with the experimental ones \cite{clement2007}.
In our calculations, the  $0^+_1$ ground states for
both $^{74,76}$Kr  are mainly oblate in nature
while the $0^+_2$ states  are predominantly
prolate and spherical, respectively (see, Tables~\ref{tab:config} and \ref{tab:frac}).
In the case of $^{74}$Kr, our calculations suggest rather large inter-band $B(E2)$
transitions between the lowest-spin states of the ground-state and the
first excited bands. This is confirmed by the  strong  
$B(E2; 0^+_2\rightarrow 2^+_1)$ and $B(E2; 2^+_2\rightarrow 2^+_1)$
values which are, about the same order of magnitude as the 
$B(E2; 2^+_1\rightarrow 0^+_1)$  rate. Note, however, that
the predicted
$B(E2; 0^+_2\rightarrow 2^+_1)$ value 
accounts for only half the experimental one  \cite{clement2007}. 
As already mentioned above, this disagreement suggests that a much 
stronger mixing between those states would be necessary to reproduce the
 large experimental $B(E2; 0^+_2\rightarrow 2^+_1)$ transition probability. 
Previous five-dimensional collective Hamiltonian (5DCH) calculations 
\cite{fu2013}
based on the relativistic PC-PK1 EDF also underestimate the
strong $B(E2; 0^+_2\rightarrow 2^+_1)$ rate for $^{74}$Kr. On the other hand, the 
5DCH calculations based on the Gogny-D1S EDF 
\cite{clement2007} account for it. 
Our results also suggest that in the case of 
$^{74}$Kr the quasi-$\gamma$
band is built on the $2^+_3$ state. The computed 
$0^+_3$ excitation energy agrees well with the experimental 
result whereas the $B(E2; 0^+_3\rightarrow
2^+_1)$ value is too small compared to the latter.

As shown in the lower panel of Fig.~\ref{fig:kr7476}, our calculations
provide  a reasonable agreement with the experimental 
data for  $^{76}$Kr. However, as in the case of 
$^{74}$Kr, they underestimate the $B(E2; 0^+_2\rightarrow 2^+_1)$ transition strength. 
Note, that the theoretical $3^+$ and $4^+$ levels in the quasi-$\gamma$
band, i.e., the second excited band built on the $2^+_3$ state, of
$^{76}$Kr are reversed. This might be a consequence of the strong level
repulsion among the $4^+$ states due to configuration mixing.

In Fig.~\ref{fig:kr9698}, we have plotted the low-energy  excitation
spectra for the neutron-rich nuclei $^{96,98}$Kr. 
which exhibit spectacular coexistence between prolate and oblate shapes
(see, Fig.~\ref{fig:pes-hfb}). As can be seen from 
Tables~\ref{tab:config} and \ref{tab:frac}, for both nuclei the $0^+_1$ and 
$0^+_2$ states are mainly arising from the oblate and
prolate configurations, respectively, while the two configurations are
more strongly mixed in $^{98}$Kr than in $^{96}$Kr. The predicted level schemes 
for both nuclei look rather similar and reproduce 
the experimental systematics \cite{albers2013,dudouet2017,flavigny2017}
for the lowest-lying states.

Note, that the level of accuracy of our results in describing
the low-energy spectra shown in
Figs.~\ref{fig:kr7476} and \ref{fig:kr9698} is, comparable with that of
the recent symmetry-projected GCM calculation, based on the 
Gogny-D1S EDF, in which  the triaxial 
deformation was included as a generating coordinate \cite{trodriguez2014}. 
In Ref.~\cite{trodriguez2014}, similar low-energy band structure to ours
has been obtained for $^{74,76}$Kr as well as for $^{96,98}$Kr.  
The energy spectra for $^{96,98}$Kr in the present calculation, however,
generally look more stretched than those obtained in Ref.~\cite{trodriguez2014}.


\subsection{Sensitivity test\label{sec:uncertainty}}



\begin{figure}[htb!]
\begin{center}
\includegraphics[width=\columnwidth]{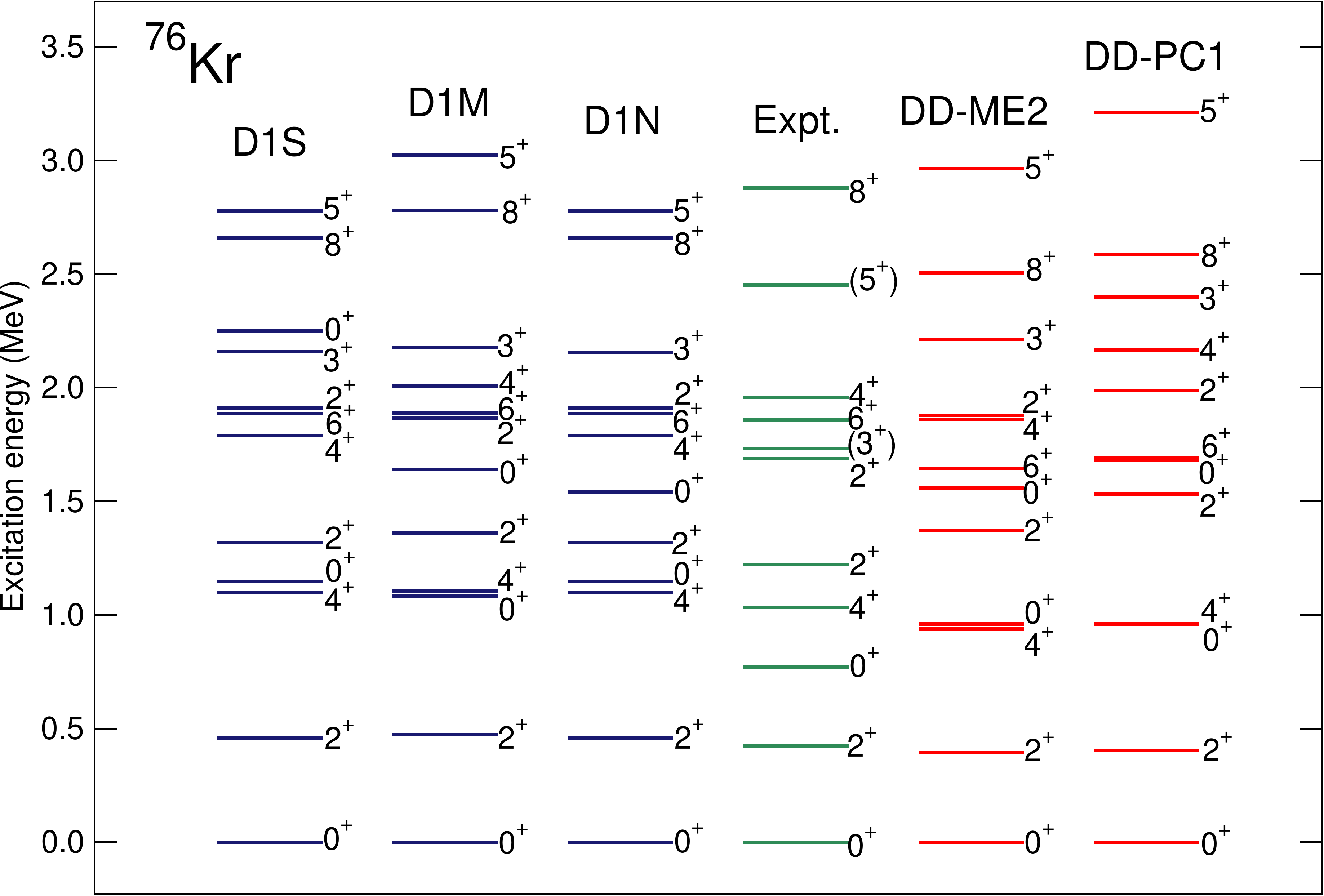} \\
\includegraphics[width=\columnwidth]{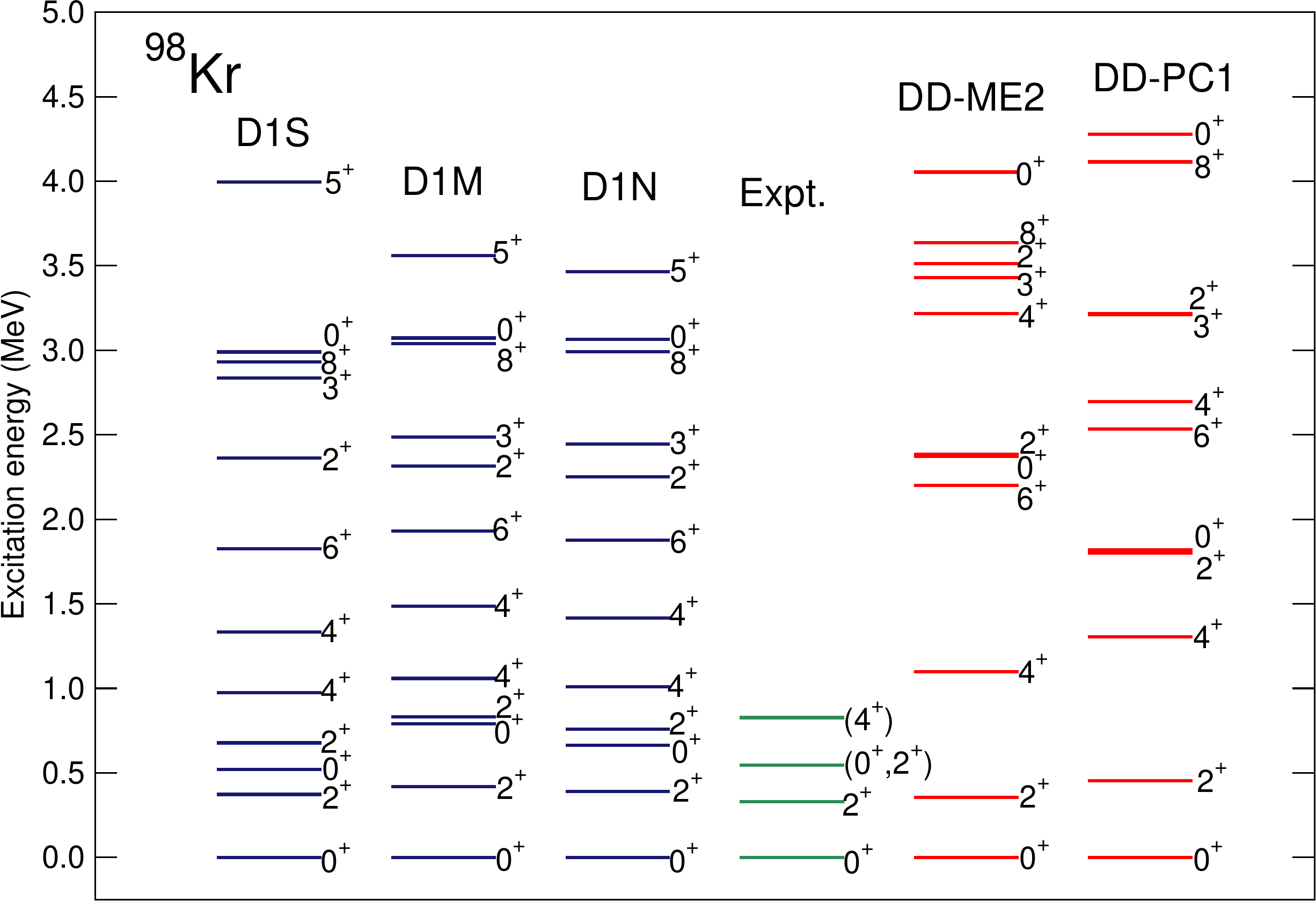}
\caption{(Color online) Comparison of the low-energy excitation spectra 
obtained for $^{76}$Kr and $^{98}$Kr with the Gogny D1S, D1M
 and D1N EDFs as well as with   the relativistic DD-ME2 and DD-PC1 EDFs. The
 corresponding experimental spectra are also included in the plot.} 
\label{fig:comp-edfs}
\end{center}
\end{figure}

Among the various factors that could affect the spectroscopic 
properties obtained for the studied nuclei, the choice of the EDF at 
the mean-field level is a relevant one since the parameters of the IBM 
Hamiltonian are determined as so to reproduce the topology of the SCMF 
energy surfaces. In this section, we analyze the sensitivity of the 
calculated excitation spectra with respect to the choice of the 
underlying EDF. To this end, in Fig.~\ref{fig:comp-edfs}, we have 
compared the low-energy excitation spectra obtained for $^{76}$Kr 
(upper panel) and $^{98}$Kr (lower panel). Calculations have been 
carried out with three different parametrizations of the Gogny-EDF, 
i.e., D1S \cite{D1S}, D1M \cite{D1M} and D1N \cite{D1N} as well as with 
two parametrizations of the relativistic mean-field Lagrangian, i.e., 
DD-ME2 \cite{lalazissis2005} and DD-PC1 \cite{DDPC1}. The  experimental 
data are also included in the  plots. As can be seen from the figure, 
all the EDFs provide similar excitation spectra for $^{76}$Kr. On the 
other hand, in the case of $^{98}$Kr, the spectra obtained with the 
three Gogny EDFs are rather similar while there are significant 
differences with the ones provided by the DD-ME2 and DD-PC1 parameter 
sets which provide much more stretched energy levels, particularly for 
the non-yrast states.


\section{Summary and perspectives\label{sec:summary}}


In this paper, we have studied the shape transition and shape 
coexistence phenomena  along the Kr isotopic chain. To this end, the 
nuclei $^{70-100}$Kr have been taken as an illustrative sample. We have 
resorted to a fermion-to-boson mapping procedure based on mapping the 
fermionic $(\beta,\gamma)$ energy contour plot onto the expectation 
value of the IBM Hamiltonian that includes configuration mixing. The 
parameters of the IBM Hamiltonian have been determined through this 
procedure and used to compute spectroscopic properties that 
characterize the structural evolution along the Kr isotopic chain. The 
microscopic input to our calculations is provided by SCMF calculations 
based on the nonrelativistic Gogny-EDF as well as different 
parametrizations of the relativistic mean-field Lagrangian. In 
particular, for the former we have considered the three parameter sets 
D1S, D1N and D1M while for the latter we have considered the DD-ME2 and 
DD-PC1 parametrizations.

The Gogny-D1M energy surfaces suggest an oblate ground state for 
$^{70}$Kr, coexisting oblate and prolate minima in the case of 
$^{72,74}$Kr and spherical-oblate-prolate triple shape coexistence for 
$^{76,78}$Kr. On the other hand, nearly spherical ground states are 
found for $^{80-86}$Kr while $\gamma$-softness emerges in the case of
 $^{88,90,92}$Kr. An oblate ground state is predicted for 
$^{94}$Kr. Moreover, prolate-oblate shape coexistence 
is obtained for the heavier nuclei
$^{96,98,100}$Kr.

The evolution of the  low-energy excitation spectra, $B(E2)$ transition 
rates and the $\rho^{2}(E0)$ values, as functions of the neutron 
number,  correlates  well with the systematics of the Gogny-D1M energy 
surfaces. Despite the simplicity of the considered (mapped) IBM 
approach, the predicted spectroscopic properties exhibit a reasonable 
agreement with the available experimental data. We have also studied 
the robustness of our approach by comparing the excitation spectra 
obtained from several nonrelativistic and relativistic EDFs. Such a 
comparison reveals no essential difference between the predictions 
obtained for neutron-deficient Kr isotopes. On the other hand, we have 
found that the relativistic and non-relativistic EDFs provide notably 
different predictions in the case of neutron-rich systems.

Several approximations have been made  at various levels of the mapping 
procedure. This leads to a  disagreement with the  experimental data  
in some spectroscopic properties. For example, near $N=40$ our approach  
does not reproduce the inter-band $B(E2; 0^+_2\rightarrow 2^+_1)$ and 
$B(E2; 2^+_2\rightarrow 2^+_1)$ transitions. Therefore, further 
improvement of our mapping procedure is still required to properly 
account for shape mixing. A similar conclusion has been reached in our 
previous studies of nuclei in this region of the nuclear chart 
\cite{nomura2016zr,nomura2017ge} regardless of the underlying EDF 
employed in the mapping procedure. A possible improvement would be to 
use a more general form of the IBM Hamiltonian that includes additional 
degrees of freedom like proton and neutron bosons. This, however, would 
increase the number of parameters in our model. Those parameters could 
not be uniquely determined just by looking at the (static) mean-field 
energy surfaces and additional microscopic input would be needed for 
the mapping procedure.

Another example is the assumption on the IBM configuration space. 
We have associated the particle-hole configurations with the mean-field
minima with larger $\beta$ deformation. 
This assumption, however, may not obviously provide a proper 
interpretation of coexisting shapes in the considered Kr nuclei. 
In addition, while we have considered only the
proton particle-hole excitations across the $Z=28$ shell gap as the  
source of shape coexistence, there could be some other
possibilities of particle-hole excitations, e.g., those of neutrons
especially around the sub-shell closure $N=40$. 
Therefore, in order to associate a configuration to each mean-field minimum in a more
unambiguous manner, a more elaborate analysis would be required to
examine the nature of the underlying SCMF 
state at each minimum to see which particle-hole components play a role,
and then incorporate it to the corresponding boson subspace. 
Such an analysis would require 
another extensive study with further complications arising, e.g., from the
inclusion of additional boson degrees of freedom, which is well beyond
the scope of the present work. Work along these lines represents an
important step for further developing our
mapping procedure and will be considered in future studies.

\acknowledgments
K.N. acknowledges support from the Japan 
Society for the Promotion of Science. This work has been supported in 
part by the QuantiXLie Centre of Excellence. The  work of LMR was 
supported by Spanish Ministry of Economy and Competitiveness (MINECO) Grants No. FPA2015-65929-P and
No. FIS2015-63770-P.

\bibliography{refs}

\begin{thebibliography}{55}%
\makeatletter
\providecommand \@ifxundefined [1]{%
 \@ifx{#1\undefined}
}%
\providecommand \@ifnum [1]{%
 \ifnum #1\expandafter \@firstoftwo
 \else \expandafter \@secondoftwo
 \fi
}%
\providecommand \@ifx [1]{%
 \ifx #1\expandafter \@firstoftwo
 \else \expandafter \@secondoftwo
 \fi
}%
\providecommand \natexlab [1]{#1}%
\providecommand \enquote  [1]{``#1''}%
\providecommand \bibnamefont  [1]{#1}%
\providecommand \bibfnamefont [1]{#1}%
\providecommand \citenamefont [1]{#1}%
\providecommand \href@noop [0]{\@secondoftwo}%
\providecommand \href [0]{\begingroup \@sanitize@url \@href}%
\providecommand \@href[1]{\@@startlink{#1}\@@href}%
\providecommand \@@href[1]{\endgroup#1\@@endlink}%
\providecommand \@sanitize@url [0]{\catcode `\\12\catcode `\$12\catcode
  `\&12\catcode `\#12\catcode `\^12\catcode `\_12\catcode `\%12\relax}%
\providecommand \@@startlink[1]{}%
\providecommand \@@endlink[0]{}%
\providecommand \url  [0]{\begingroup\@sanitize@url \@url }%
\providecommand \@url [1]{\endgroup\@href {#1}{\urlprefix }}%
\providecommand \urlprefix  [0]{URL }%
\providecommand \Eprint [0]{\href }%
\providecommand \doibase [0]{http://dx.doi.org/}%
\providecommand \selectlanguage [0]{\@gobble}%
\providecommand \bibinfo  [0]{\@secondoftwo}%
\providecommand \bibfield  [0]{\@secondoftwo}%
\providecommand \translation [1]{[#1]}%
\providecommand \BibitemOpen [0]{}%
\providecommand \bibitemStop [0]{}%
\providecommand \bibitemNoStop [0]{.\EOS\space}%
\providecommand \EOS [0]{\spacefactor3000\relax}%
\providecommand \BibitemShut  [1]{\csname bibitem#1\endcsname}%
\let\auto@bib@innerbib\@empty
\bibitem [{\citenamefont {Casten}(2005)}]{CasBook}%
  \BibitemOpen
  \bibfield  {author} {\bibinfo {author} {\bibfnamefont {R.~F.}\ \bibnamefont
  {Casten}},\ }\href@noop {} {\emph {\bibinfo {title} {Nuclear Structure from a
  Simple Perspective}}}\ (\bibinfo  {publisher} {Oxford University Press,
  Oxford, England},\ \bibinfo {year} {2005})\BibitemShut {NoStop}%
\bibitem [{\citenamefont {Cejnar}\ \emph {et~al.}(2010)\citenamefont {Cejnar},
  \citenamefont {Jolie},\ and\ \citenamefont {Casten}}]{cejnar2010}%
  \BibitemOpen
  \bibfield  {author} {\bibinfo {author} {\bibfnamefont {P.}~\bibnamefont
  {Cejnar}}, \bibinfo {author} {\bibfnamefont {J.}~\bibnamefont {Jolie}}, \
  and\ \bibinfo {author} {\bibfnamefont {R.~F.}\ \bibnamefont {Casten}},\
  }\href {\doibase 10.1103/RevModPhys.82.2155} {\bibfield  {journal} {\bibinfo
  {journal} {Rev. Mod. Phys.}\ }\textbf {\bibinfo {volume} {82}},\ \bibinfo
  {pages} {2155} (\bibinfo {year} {2010})}\BibitemShut {NoStop}%
\bibitem [{\citenamefont {Heyde}\ and\ \citenamefont {Wood}(2011)}]{heyde2011}%
  \BibitemOpen
  \bibfield  {author} {\bibinfo {author} {\bibfnamefont {K.}~\bibnamefont
  {Heyde}}\ and\ \bibinfo {author} {\bibfnamefont {J.~L.}\ \bibnamefont
  {Wood}},\ }\href {\doibase 10.1103/RevModPhys.83.1467} {\bibfield  {journal}
  {\bibinfo  {journal} {Rev. Mod. Phys.}\ }\textbf {\bibinfo {volume} {83}},\
  \bibinfo {pages} {1467} (\bibinfo {year} {2011})}\BibitemShut {NoStop}%
\bibitem [{\citenamefont {Cl\'ement}\ \emph {et~al.}(2007)\citenamefont
  {Cl\'ement}, \citenamefont {G\"orgen}, \citenamefont {Korten}, \citenamefont
  {Bouchez}, \citenamefont {Chatillon}, \citenamefont {Delaroche},
  \citenamefont {Girod}, \citenamefont {Goutte}, \citenamefont {H\"urstel},
  \citenamefont {Coz}, \citenamefont {Obertelli}, \citenamefont {P\'eru},
  \citenamefont {Theisen}, \citenamefont {Wilson}, \citenamefont
  {Zieli\ifmmode~\acute{n}\else \'{n}\fi{}ska}, \citenamefont {Andreoiu},
  \citenamefont {Becker}, \citenamefont {Butler}, \citenamefont {Casandjian},
  \citenamefont {Catford}, \citenamefont {Czosnyka}, \citenamefont {France},
  \citenamefont {Gerl}, \citenamefont {Herzberg}, \citenamefont {Iwanicki},
  \citenamefont {Jenkins}, \citenamefont {Jones}, \citenamefont {Napiorkowski},
  \citenamefont {Sletten},\ and\ \citenamefont {Timis}}]{clement2007}%
  \BibitemOpen
  \bibfield  {author} {\bibinfo {author} {\bibfnamefont {E.}~\bibnamefont
  {Cl\'ement}}, \bibinfo {author} {\bibfnamefont {A.}~\bibnamefont {G\"orgen}},
  \bibinfo {author} {\bibfnamefont {W.}~\bibnamefont {Korten}}, \bibinfo
  {author} {\bibfnamefont {E.}~\bibnamefont {Bouchez}}, \bibinfo {author}
  {\bibfnamefont {A.}~\bibnamefont {Chatillon}}, \bibinfo {author}
  {\bibfnamefont {J.-P.}\ \bibnamefont {Delaroche}}, \bibinfo {author}
  {\bibfnamefont {M.}~\bibnamefont {Girod}}, \bibinfo {author} {\bibfnamefont
  {H.}~\bibnamefont {Goutte}}, \bibinfo {author} {\bibfnamefont
  {A.}~\bibnamefont {H\"urstel}}, \bibinfo {author} {\bibfnamefont {Y.~L.}\
  \bibnamefont {Coz}}, \bibinfo {author} {\bibfnamefont {A.}~\bibnamefont
  {Obertelli}}, \bibinfo {author} {\bibfnamefont {S.}~\bibnamefont {P\'eru}},
  \bibinfo {author} {\bibfnamefont {C.}~\bibnamefont {Theisen}}, \bibinfo
  {author} {\bibfnamefont {J.~N.}\ \bibnamefont {Wilson}}, \bibinfo {author}
  {\bibfnamefont {M.}~\bibnamefont {Zieli\ifmmode~\acute{n}\else
  \'{n}\fi{}ska}}, \bibinfo {author} {\bibfnamefont {C.}~\bibnamefont
  {Andreoiu}}, \bibinfo {author} {\bibfnamefont {F.}~\bibnamefont {Becker}},
  \bibinfo {author} {\bibfnamefont {P.~A.}\ \bibnamefont {Butler}}, \bibinfo
  {author} {\bibfnamefont {J.~M.}\ \bibnamefont {Casandjian}}, \bibinfo
  {author} {\bibfnamefont {W.~N.}\ \bibnamefont {Catford}}, \bibinfo {author}
  {\bibfnamefont {T.}~\bibnamefont {Czosnyka}}, \bibinfo {author}
  {\bibfnamefont {G.~d.}\ \bibnamefont {France}}, \bibinfo {author}
  {\bibfnamefont {J.}~\bibnamefont {Gerl}}, \bibinfo {author} {\bibfnamefont
  {R.-D.}\ \bibnamefont {Herzberg}}, \bibinfo {author} {\bibfnamefont
  {J.}~\bibnamefont {Iwanicki}}, \bibinfo {author} {\bibfnamefont {D.~G.}\
  \bibnamefont {Jenkins}}, \bibinfo {author} {\bibfnamefont {G.~D.}\
  \bibnamefont {Jones}}, \bibinfo {author} {\bibfnamefont {P.~J.}\ \bibnamefont
  {Napiorkowski}}, \bibinfo {author} {\bibfnamefont {G.}~\bibnamefont
  {Sletten}}, \ and\ \bibinfo {author} {\bibfnamefont {C.~N.}\ \bibnamefont
  {Timis}},\ }\href {\doibase 10.1103/PhysRevC.75.054313} {\bibfield  {journal}
  {\bibinfo  {journal} {Phys. Rev. C}\ }\textbf {\bibinfo {volume} {75}},\
  \bibinfo {pages} {054313} (\bibinfo {year} {2007})}\BibitemShut {NoStop}%
\bibitem [{\citenamefont {Bender}\ \emph {et~al.}(2006)\citenamefont {Bender},
  \citenamefont {Bonche},\ and\ \citenamefont {Heenen}}]{bender2006}%
  \BibitemOpen
  \bibfield  {author} {\bibinfo {author} {\bibfnamefont {M.}~\bibnamefont
  {Bender}}, \bibinfo {author} {\bibfnamefont {P.}~\bibnamefont {Bonche}}, \
  and\ \bibinfo {author} {\bibfnamefont {P.-H.}\ \bibnamefont {Heenen}},\
  }\href {\doibase 10.1103/PhysRevC.74.024312} {\bibfield  {journal} {\bibinfo
  {journal} {Phys. Rev. C}\ }\textbf {\bibinfo {volume} {74}},\ \bibinfo
  {pages} {024312} (\bibinfo {year} {2006})}\BibitemShut {NoStop}%
\bibitem [{\citenamefont {Fu}\ \emph {et~al.}(2013)\citenamefont {Fu},
  \citenamefont {Mei}, \citenamefont {Xiang}, \citenamefont {Li}, \citenamefont
  {Yao},\ and\ \citenamefont {Meng}}]{fu2013}%
  \BibitemOpen
  \bibfield  {author} {\bibinfo {author} {\bibfnamefont {Y.}~\bibnamefont
  {Fu}}, \bibinfo {author} {\bibfnamefont {H.}~\bibnamefont {Mei}}, \bibinfo
  {author} {\bibfnamefont {J.}~\bibnamefont {Xiang}}, \bibinfo {author}
  {\bibfnamefont {Z.~P.}\ \bibnamefont {Li}}, \bibinfo {author} {\bibfnamefont
  {J.~M.}\ \bibnamefont {Yao}}, \ and\ \bibinfo {author} {\bibfnamefont
  {J.}~\bibnamefont {Meng}},\ }\href {\doibase 10.1103/PhysRevC.87.054305}
  {\bibfield  {journal} {\bibinfo  {journal} {Phys. Rev. C}\ }\textbf {\bibinfo
  {volume} {87}},\ \bibinfo {pages} {054305} (\bibinfo {year}
  {2013})}\BibitemShut {NoStop}%
\bibitem [{\citenamefont {Rodr\'{\i}guez}(2014)}]{trodriguez2014}%
  \BibitemOpen
  \bibfield  {author} {\bibinfo {author} {\bibfnamefont {T.~R.}\ \bibnamefont
  {Rodr\'{\i}guez}},\ }\href@noop {} {\bibfield  {journal} {\bibinfo  {journal}
  {Phys. Rev. C}\ }\textbf {\bibinfo {volume} {90}},\ \bibinfo {pages} {034306}
  (\bibinfo {year} {2014})}\BibitemShut {NoStop}%
\bibitem [{\citenamefont {Petrovici}\ \emph {et~al.}(2000)\citenamefont
  {Petrovici}, \citenamefont {Schmid},\ and\ \citenamefont
  {Faessler}}]{PETROVICI2000}%
  \BibitemOpen
  \bibfield  {author} {\bibinfo {author} {\bibfnamefont {A.}~\bibnamefont
  {Petrovici}}, \bibinfo {author} {\bibfnamefont {K.}~\bibnamefont {Schmid}}, \
  and\ \bibinfo {author} {\bibfnamefont {A.}~\bibnamefont {Faessler}},\ }\href
  {\doibase http://dx.doi.org/10.1016/S0375-9474(99)00811-8} {\bibfield
  {journal} {\bibinfo  {journal} {Nuclear Physics A}\ }\textbf {\bibinfo
  {volume} {665}},\ \bibinfo {pages} {333 } (\bibinfo {year}
  {2000})}\BibitemShut {NoStop}%
\bibitem [{\citenamefont {Sato}\ and\ \citenamefont
  {Hinohara}(2011)}]{SATO2011}%
  \BibitemOpen
  \bibfield  {author} {\bibinfo {author} {\bibfnamefont {K.}~\bibnamefont
  {Sato}}\ and\ \bibinfo {author} {\bibfnamefont {N.}~\bibnamefont
  {Hinohara}},\ }\href {\doibase
  http://dx.doi.org/10.1016/j.nuclphysa.2010.11.003} {\bibfield  {journal}
  {\bibinfo  {journal} {Nuclear Physics A}\ }\textbf {\bibinfo {volume}
  {849}},\ \bibinfo {pages} {53 } (\bibinfo {year} {2011})}\BibitemShut
  {NoStop}%
\bibitem [{\citenamefont {Naimi}\ \emph {et~al.}(2010)\citenamefont {Naimi},
  \citenamefont {Audi}, \citenamefont {Beck}, \citenamefont {Blaum},
  \citenamefont {B\"ohm}, \citenamefont {Borgmann}, \citenamefont
  {Breitenfeldt}, \citenamefont {George}, \citenamefont {Herfurth},
  \citenamefont {Herlert}, \citenamefont {Kowalska}, \citenamefont {Kreim},
  \citenamefont {Lunney}, \citenamefont {Neidherr}, \citenamefont {Rosenbusch},
  \citenamefont {Schwarz}, \citenamefont {Schweikhard},\ and\ \citenamefont
  {Zuber}}]{naimi2010}%
  \BibitemOpen
  \bibfield  {author} {\bibinfo {author} {\bibfnamefont {S.}~\bibnamefont
  {Naimi}}, \bibinfo {author} {\bibfnamefont {G.}~\bibnamefont {Audi}},
  \bibinfo {author} {\bibfnamefont {D.}~\bibnamefont {Beck}}, \bibinfo {author}
  {\bibfnamefont {K.}~\bibnamefont {Blaum}}, \bibinfo {author} {\bibfnamefont
  {C.}~\bibnamefont {B\"ohm}}, \bibinfo {author} {\bibfnamefont
  {C.}~\bibnamefont {Borgmann}}, \bibinfo {author} {\bibfnamefont
  {M.}~\bibnamefont {Breitenfeldt}}, \bibinfo {author} {\bibfnamefont
  {S.}~\bibnamefont {George}}, \bibinfo {author} {\bibfnamefont
  {F.}~\bibnamefont {Herfurth}}, \bibinfo {author} {\bibfnamefont
  {A.}~\bibnamefont {Herlert}}, \bibinfo {author} {\bibfnamefont
  {M.}~\bibnamefont {Kowalska}}, \bibinfo {author} {\bibfnamefont
  {S.}~\bibnamefont {Kreim}}, \bibinfo {author} {\bibfnamefont
  {D.}~\bibnamefont {Lunney}}, \bibinfo {author} {\bibfnamefont
  {D.}~\bibnamefont {Neidherr}}, \bibinfo {author} {\bibfnamefont
  {M.}~\bibnamefont {Rosenbusch}}, \bibinfo {author} {\bibfnamefont
  {S.}~\bibnamefont {Schwarz}}, \bibinfo {author} {\bibfnamefont
  {L.}~\bibnamefont {Schweikhard}}, \ and\ \bibinfo {author} {\bibfnamefont
  {K.}~\bibnamefont {Zuber}},\ }\href {\doibase 10.1103/PhysRevLett.105.032502}
  {\bibfield  {journal} {\bibinfo  {journal} {Phys. Rev. Lett.}\ }\textbf
  {\bibinfo {volume} {105}},\ \bibinfo {pages} {032502} (\bibinfo {year}
  {2010})}\BibitemShut {NoStop}%
\bibitem [{\citenamefont {Albers}\ \emph {et~al.}(2012)\citenamefont {Albers},
  \citenamefont {Warr}, \citenamefont {Nomura}, \citenamefont {Blazhev},
  \citenamefont {Jolie}, \citenamefont {M\"ucher}, \citenamefont {Bastin},
  \citenamefont {Bauer}, \citenamefont {Bernards}, \citenamefont {Bettermann},
  \citenamefont {Bildstein}, \citenamefont {Butterworth}, \citenamefont
  {Cappellazzo}, \citenamefont {Cederk\"all}, \citenamefont {Cline},
  \citenamefont {Darby}, \citenamefont {Das~Gupta}, \citenamefont {Daugas},
  \citenamefont {Davinson}, \citenamefont {De~Witte}, \citenamefont {Diriken},
  \citenamefont {Filipescu}, \citenamefont {Fiori}, \citenamefont {Fransen},
  \citenamefont {Gaffney}, \citenamefont {Georgiev}, \citenamefont
  {Gernh\"auser}, \citenamefont {Hackstein}, \citenamefont {Heinze},
  \citenamefont {Hess}, \citenamefont {Huyse}, \citenamefont {Jenkins},
  \citenamefont {Konki}, \citenamefont {Kowalczyk}, \citenamefont {Kr\"oll},
  \citenamefont {Kr\"ucken}, \citenamefont {Litzinger}, \citenamefont {Lutter},
  \citenamefont {Marginean}, \citenamefont {Mihai}, \citenamefont {Moschner},
  \citenamefont {Napiorkowski}, \citenamefont {Nara~Singh}, \citenamefont
  {Nowak}, \citenamefont {Otsuka}, \citenamefont {Pakarinen}, \citenamefont
  {Pfeiffer}, \citenamefont {Radeck}, \citenamefont {Reiter}, \citenamefont
  {Rigby}, \citenamefont {Robledo}, \citenamefont {Rodr\'iguez-Guzm\'an},
  \citenamefont {Rudigier}, \citenamefont {Sarriguren}, \citenamefont {Scheck},
  \citenamefont {Seidlitz}, \citenamefont {Siebeck}, \citenamefont {Simpson},
  \citenamefont {Th\"ole}, \citenamefont {Thomas}, \citenamefont {Van~de
  Walle}, \citenamefont {Van~Duppen}, \citenamefont {Vermeulen}, \citenamefont
  {Voulot}, \citenamefont {Wadsworth}, \citenamefont {Wenander}, \citenamefont
  {Wimmer}, \citenamefont {Zell},\ and\ \citenamefont
  {Zielinska}}]{albers2012}%
  \BibitemOpen
  \bibfield  {author} {\bibinfo {author} {\bibfnamefont {M.}~\bibnamefont
  {Albers}}, \bibinfo {author} {\bibfnamefont {N.}~\bibnamefont {Warr}},
  \bibinfo {author} {\bibfnamefont {K.}~\bibnamefont {Nomura}}, \bibinfo
  {author} {\bibfnamefont {A.}~\bibnamefont {Blazhev}}, \bibinfo {author}
  {\bibfnamefont {J.}~\bibnamefont {Jolie}}, \bibinfo {author} {\bibfnamefont
  {D.}~\bibnamefont {M\"ucher}}, \bibinfo {author} {\bibfnamefont
  {B.}~\bibnamefont {Bastin}}, \bibinfo {author} {\bibfnamefont
  {C.}~\bibnamefont {Bauer}}, \bibinfo {author} {\bibfnamefont
  {C.}~\bibnamefont {Bernards}}, \bibinfo {author} {\bibfnamefont
  {L.}~\bibnamefont {Bettermann}}, \bibinfo {author} {\bibfnamefont
  {V.}~\bibnamefont {Bildstein}}, \bibinfo {author} {\bibfnamefont
  {J.}~\bibnamefont {Butterworth}}, \bibinfo {author} {\bibfnamefont
  {M.}~\bibnamefont {Cappellazzo}}, \bibinfo {author} {\bibfnamefont
  {J.}~\bibnamefont {Cederk\"all}}, \bibinfo {author} {\bibfnamefont
  {D.}~\bibnamefont {Cline}}, \bibinfo {author} {\bibfnamefont
  {I.}~\bibnamefont {Darby}}, \bibinfo {author} {\bibfnamefont
  {S.}~\bibnamefont {Das~Gupta}}, \bibinfo {author} {\bibfnamefont {J.~M.}\
  \bibnamefont {Daugas}}, \bibinfo {author} {\bibfnamefont {T.}~\bibnamefont
  {Davinson}}, \bibinfo {author} {\bibfnamefont {H.}~\bibnamefont {De~Witte}},
  \bibinfo {author} {\bibfnamefont {J.}~\bibnamefont {Diriken}}, \bibinfo
  {author} {\bibfnamefont {D.}~\bibnamefont {Filipescu}}, \bibinfo {author}
  {\bibfnamefont {E.}~\bibnamefont {Fiori}}, \bibinfo {author} {\bibfnamefont
  {C.}~\bibnamefont {Fransen}}, \bibinfo {author} {\bibfnamefont {L.~P.}\
  \bibnamefont {Gaffney}}, \bibinfo {author} {\bibfnamefont {G.}~\bibnamefont
  {Georgiev}}, \bibinfo {author} {\bibfnamefont {R.}~\bibnamefont
  {Gernh\"auser}}, \bibinfo {author} {\bibfnamefont {M.}~\bibnamefont
  {Hackstein}}, \bibinfo {author} {\bibfnamefont {S.}~\bibnamefont {Heinze}},
  \bibinfo {author} {\bibfnamefont {H.}~\bibnamefont {Hess}}, \bibinfo {author}
  {\bibfnamefont {M.}~\bibnamefont {Huyse}}, \bibinfo {author} {\bibfnamefont
  {D.}~\bibnamefont {Jenkins}}, \bibinfo {author} {\bibfnamefont
  {J.}~\bibnamefont {Konki}}, \bibinfo {author} {\bibfnamefont
  {M.}~\bibnamefont {Kowalczyk}}, \bibinfo {author} {\bibfnamefont
  {T.}~\bibnamefont {Kr\"oll}}, \bibinfo {author} {\bibfnamefont
  {R.}~\bibnamefont {Kr\"ucken}}, \bibinfo {author} {\bibfnamefont
  {J.}~\bibnamefont {Litzinger}}, \bibinfo {author} {\bibfnamefont
  {R.}~\bibnamefont {Lutter}}, \bibinfo {author} {\bibfnamefont
  {N.}~\bibnamefont {Marginean}}, \bibinfo {author} {\bibfnamefont
  {C.}~\bibnamefont {Mihai}}, \bibinfo {author} {\bibfnamefont
  {K.}~\bibnamefont {Moschner}}, \bibinfo {author} {\bibfnamefont
  {P.}~\bibnamefont {Napiorkowski}}, \bibinfo {author} {\bibfnamefont {B.~S.}\
  \bibnamefont {Nara~Singh}}, \bibinfo {author} {\bibfnamefont
  {K.}~\bibnamefont {Nowak}}, \bibinfo {author} {\bibfnamefont
  {T.}~\bibnamefont {Otsuka}}, \bibinfo {author} {\bibfnamefont
  {J.}~\bibnamefont {Pakarinen}}, \bibinfo {author} {\bibfnamefont
  {M.}~\bibnamefont {Pfeiffer}}, \bibinfo {author} {\bibfnamefont
  {D.}~\bibnamefont {Radeck}}, \bibinfo {author} {\bibfnamefont
  {P.}~\bibnamefont {Reiter}}, \bibinfo {author} {\bibfnamefont
  {S.}~\bibnamefont {Rigby}}, \bibinfo {author} {\bibfnamefont {L.~M.}\
  \bibnamefont {Robledo}}, \bibinfo {author} {\bibfnamefont {R.}~\bibnamefont
  {Rodr\'iguez-Guzm\'an}}, \bibinfo {author} {\bibfnamefont {M.}~\bibnamefont
  {Rudigier}}, \bibinfo {author} {\bibfnamefont {P.}~\bibnamefont
  {Sarriguren}}, \bibinfo {author} {\bibfnamefont {M.}~\bibnamefont {Scheck}},
  \bibinfo {author} {\bibfnamefont {M.}~\bibnamefont {Seidlitz}}, \bibinfo
  {author} {\bibfnamefont {B.}~\bibnamefont {Siebeck}}, \bibinfo {author}
  {\bibfnamefont {G.}~\bibnamefont {Simpson}}, \bibinfo {author} {\bibfnamefont
  {P.}~\bibnamefont {Th\"ole}}, \bibinfo {author} {\bibfnamefont
  {T.}~\bibnamefont {Thomas}}, \bibinfo {author} {\bibfnamefont
  {J.}~\bibnamefont {Van~de Walle}}, \bibinfo {author} {\bibfnamefont
  {P.}~\bibnamefont {Van~Duppen}}, \bibinfo {author} {\bibfnamefont
  {M.}~\bibnamefont {Vermeulen}}, \bibinfo {author} {\bibfnamefont
  {D.}~\bibnamefont {Voulot}}, \bibinfo {author} {\bibfnamefont
  {R.}~\bibnamefont {Wadsworth}}, \bibinfo {author} {\bibfnamefont
  {F.}~\bibnamefont {Wenander}}, \bibinfo {author} {\bibfnamefont
  {K.}~\bibnamefont {Wimmer}}, \bibinfo {author} {\bibfnamefont {K.~O.}\
  \bibnamefont {Zell}}, \ and\ \bibinfo {author} {\bibfnamefont
  {M.}~\bibnamefont {Zielinska}},\ }\href {\doibase
  10.1103/PhysRevLett.108.062701} {\bibfield  {journal} {\bibinfo  {journal}
  {Phys. Rev. Lett.}\ }\textbf {\bibinfo {volume} {108}},\ \bibinfo {pages}
  {062701} (\bibinfo {year} {2012})}\BibitemShut {NoStop}%
\bibitem [{\citenamefont {Albers}\ \emph {et~al.}(2013)\citenamefont {Albers},
  \citenamefont {Nomura}, \citenamefont {Warr}, \citenamefont {Blazhev},
  \citenamefont {Jolie}, \citenamefont {M^^c3^^bccher}, \citenamefont {Bastin},
  \citenamefont {Bauer}, \citenamefont {Bernards}, \citenamefont {Bettermann},
  \citenamefont {Bildstein}, \citenamefont {Butterworth}, \citenamefont
  {Cappellazzo}, \citenamefont {Cederk^^c3^^a4ll}, \citenamefont {Cline},
  \citenamefont {Darby}, \citenamefont {Gupta}, \citenamefont {Daugas},
  \citenamefont {Davinson}, \citenamefont {Witte}, \citenamefont {Diriken},
  \citenamefont {Filipescu}, \citenamefont {Fiori}, \citenamefont {Fransen},
  \citenamefont {Gaffney}, \citenamefont {Georgiev}, \citenamefont
  {Gernh^^c3^^a4user}, \citenamefont {Hackstein}, \citenamefont {Heinze},
  \citenamefont {Hess}, \citenamefont {Huyse}, \citenamefont {Jenkins},
  \citenamefont {Konki}, \citenamefont {Kowalczyk}, \citenamefont
  {Kr^^c3^^b6ll}, \citenamefont {Kr^^c3^^bccken}, \citenamefont {Litzinger},
  \citenamefont {Lutter}, \citenamefont {Marginean}, \citenamefont {Mihai},
  \citenamefont {Moschner}, \citenamefont {Napiorkowski}, \citenamefont
  {Singh}, \citenamefont {Nowak}, \citenamefont {Pakarinen}, \citenamefont
  {Pfeiffer}, \citenamefont {Radeck}, \citenamefont {Reiter}, \citenamefont
  {Rigby}, \citenamefont {Robledo}, \citenamefont
  {Rodr^^c3^^adguez-Guzm^^c3^^a1n}, \citenamefont {Rudigier}, \citenamefont
  {Scheck}, \citenamefont {Seidlitz}, \citenamefont {Siebeck}, \citenamefont
  {Simpson}, \citenamefont {Th^^c3^^b6le}, \citenamefont {Thomas},
  \citenamefont {de~Walle}, \citenamefont {Duppen}, \citenamefont {Vermeulen},
  \citenamefont {Voulot}, \citenamefont {Wadsworth}, \citenamefont {Wenander},
  \citenamefont {Wimmer}, \citenamefont {Zell},\ and\ \citenamefont
  {Zielinska}}]{albers2013}%
  \BibitemOpen
  \bibfield  {author} {\bibinfo {author} {\bibfnamefont {M.}~\bibnamefont
  {Albers}}, \bibinfo {author} {\bibfnamefont {K.}~\bibnamefont {Nomura}},
  \bibinfo {author} {\bibfnamefont {N.}~\bibnamefont {Warr}}, \bibinfo {author}
  {\bibfnamefont {A.}~\bibnamefont {Blazhev}}, \bibinfo {author} {\bibfnamefont
  {J.}~\bibnamefont {Jolie}}, \bibinfo {author} {\bibfnamefont
  {D.}~\bibnamefont {M^^c3^^bccher}}, \bibinfo {author} {\bibfnamefont
  {B.}~\bibnamefont {Bastin}}, \bibinfo {author} {\bibfnamefont
  {C.}~\bibnamefont {Bauer}}, \bibinfo {author} {\bibfnamefont
  {C.}~\bibnamefont {Bernards}}, \bibinfo {author} {\bibfnamefont
  {L.}~\bibnamefont {Bettermann}}, \bibinfo {author} {\bibfnamefont
  {V.}~\bibnamefont {Bildstein}}, \bibinfo {author} {\bibfnamefont
  {J.}~\bibnamefont {Butterworth}}, \bibinfo {author} {\bibfnamefont
  {M.}~\bibnamefont {Cappellazzo}}, \bibinfo {author} {\bibfnamefont
  {J.}~\bibnamefont {Cederk^^c3^^a4ll}}, \bibinfo {author} {\bibfnamefont
  {D.}~\bibnamefont {Cline}}, \bibinfo {author} {\bibfnamefont
  {I.}~\bibnamefont {Darby}}, \bibinfo {author} {\bibfnamefont {S.~D.}\
  \bibnamefont {Gupta}}, \bibinfo {author} {\bibfnamefont {J.}~\bibnamefont
  {Daugas}}, \bibinfo {author} {\bibfnamefont {T.}~\bibnamefont {Davinson}},
  \bibinfo {author} {\bibfnamefont {H.~D.}\ \bibnamefont {Witte}}, \bibinfo
  {author} {\bibfnamefont {J.}~\bibnamefont {Diriken}}, \bibinfo {author}
  {\bibfnamefont {D.}~\bibnamefont {Filipescu}}, \bibinfo {author}
  {\bibfnamefont {E.}~\bibnamefont {Fiori}}, \bibinfo {author} {\bibfnamefont
  {C.}~\bibnamefont {Fransen}}, \bibinfo {author} {\bibfnamefont
  {L.}~\bibnamefont {Gaffney}}, \bibinfo {author} {\bibfnamefont
  {G.}~\bibnamefont {Georgiev}}, \bibinfo {author} {\bibfnamefont
  {R.}~\bibnamefont {Gernh^^c3^^a4user}}, \bibinfo {author} {\bibfnamefont
  {M.}~\bibnamefont {Hackstein}}, \bibinfo {author} {\bibfnamefont
  {S.}~\bibnamefont {Heinze}}, \bibinfo {author} {\bibfnamefont
  {H.}~\bibnamefont {Hess}}, \bibinfo {author} {\bibfnamefont {M.}~\bibnamefont
  {Huyse}}, \bibinfo {author} {\bibfnamefont {D.}~\bibnamefont {Jenkins}},
  \bibinfo {author} {\bibfnamefont {J.}~\bibnamefont {Konki}}, \bibinfo
  {author} {\bibfnamefont {M.}~\bibnamefont {Kowalczyk}}, \bibinfo {author}
  {\bibfnamefont {T.}~\bibnamefont {Kr^^c3^^b6ll}}, \bibinfo {author}
  {\bibfnamefont {R.}~\bibnamefont {Kr^^c3^^bccken}}, \bibinfo {author}
  {\bibfnamefont {J.}~\bibnamefont {Litzinger}}, \bibinfo {author}
  {\bibfnamefont {R.}~\bibnamefont {Lutter}}, \bibinfo {author} {\bibfnamefont
  {N.}~\bibnamefont {Marginean}}, \bibinfo {author} {\bibfnamefont
  {C.}~\bibnamefont {Mihai}}, \bibinfo {author} {\bibfnamefont
  {K.}~\bibnamefont {Moschner}}, \bibinfo {author} {\bibfnamefont
  {P.}~\bibnamefont {Napiorkowski}}, \bibinfo {author} {\bibfnamefont {B.~N.}\
  \bibnamefont {Singh}}, \bibinfo {author} {\bibfnamefont {K.}~\bibnamefont
  {Nowak}}, \bibinfo {author} {\bibfnamefont {J.}~\bibnamefont {Pakarinen}},
  \bibinfo {author} {\bibfnamefont {M.}~\bibnamefont {Pfeiffer}}, \bibinfo
  {author} {\bibfnamefont {D.}~\bibnamefont {Radeck}}, \bibinfo {author}
  {\bibfnamefont {P.}~\bibnamefont {Reiter}}, \bibinfo {author} {\bibfnamefont
  {S.}~\bibnamefont {Rigby}}, \bibinfo {author} {\bibfnamefont
  {L.}~\bibnamefont {Robledo}}, \bibinfo {author} {\bibfnamefont
  {R.}~\bibnamefont {Rodr^^c3^^adguez-Guzm^^c3^^a1n}}, \bibinfo {author}
  {\bibfnamefont {M.}~\bibnamefont {Rudigier}}, \bibinfo {author}
  {\bibfnamefont {M.}~\bibnamefont {Scheck}}, \bibinfo {author} {\bibfnamefont
  {M.}~\bibnamefont {Seidlitz}}, \bibinfo {author} {\bibfnamefont
  {B.}~\bibnamefont {Siebeck}}, \bibinfo {author} {\bibfnamefont
  {G.}~\bibnamefont {Simpson}}, \bibinfo {author} {\bibfnamefont
  {P.}~\bibnamefont {Th^^c3^^b6le}}, \bibinfo {author} {\bibfnamefont
  {T.}~\bibnamefont {Thomas}}, \bibinfo {author} {\bibfnamefont {J.~V.}\
  \bibnamefont {de~Walle}}, \bibinfo {author} {\bibfnamefont {P.~V.}\
  \bibnamefont {Duppen}}, \bibinfo {author} {\bibfnamefont {M.}~\bibnamefont
  {Vermeulen}}, \bibinfo {author} {\bibfnamefont {D.}~\bibnamefont {Voulot}},
  \bibinfo {author} {\bibfnamefont {R.}~\bibnamefont {Wadsworth}}, \bibinfo
  {author} {\bibfnamefont {F.}~\bibnamefont {Wenander}}, \bibinfo {author}
  {\bibfnamefont {K.}~\bibnamefont {Wimmer}}, \bibinfo {author} {\bibfnamefont
  {K.}~\bibnamefont {Zell}}, \ and\ \bibinfo {author} {\bibfnamefont
  {M.}~\bibnamefont {Zielinska}},\ }\href {\doibase
  http://dx.doi.org/10.1016/j.nuclphysa.2013.01.013} {\bibfield  {journal}
  {\bibinfo  {journal} {Nuclear Physics A}\ }\textbf {\bibinfo {volume}
  {899}},\ \bibinfo {pages} {1 } (\bibinfo {year} {2013})}\BibitemShut
  {NoStop}%
\bibitem [{\citenamefont {Rzaca-Urban}\ \emph {et~al.}(2017)\citenamefont
  {Rzaca-Urban}, \citenamefont {Sieja}, \citenamefont {Urban}, \citenamefont
  {Czerwi\ifmmode~\acute{n}\else \'{n}\fi{}ski}, \citenamefont {Blanc},
  \citenamefont {Jentschel}, \citenamefont {Mutti}, \citenamefont {K\"oster},
  \citenamefont {Soldner}, \citenamefont {de~France}, \citenamefont {Simpson},\
  and\ \citenamefont {Ur}}]{rzkacaurban2017}%
  \BibitemOpen
  \bibfield  {author} {\bibinfo {author} {\bibfnamefont {T.}~\bibnamefont
  {Rzaca-Urban}}, \bibinfo {author} {\bibfnamefont {K.}~\bibnamefont {Sieja}},
  \bibinfo {author} {\bibfnamefont {W.}~\bibnamefont {Urban}}, \bibinfo
  {author} {\bibfnamefont {M.}~\bibnamefont {Czerwi\ifmmode~\acute{n}\else
  \'{n}\fi{}ski}}, \bibinfo {author} {\bibfnamefont {A.}~\bibnamefont {Blanc}},
  \bibinfo {author} {\bibfnamefont {M.}~\bibnamefont {Jentschel}}, \bibinfo
  {author} {\bibfnamefont {P.}~\bibnamefont {Mutti}}, \bibinfo {author}
  {\bibfnamefont {U.}~\bibnamefont {K\"oster}}, \bibinfo {author}
  {\bibfnamefont {T.}~\bibnamefont {Soldner}}, \bibinfo {author} {\bibfnamefont
  {G.}~\bibnamefont {de~France}}, \bibinfo {author} {\bibfnamefont {G.~S.}\
  \bibnamefont {Simpson}}, \ and\ \bibinfo {author} {\bibfnamefont {C.~A.}\
  \bibnamefont {Ur}},\ }\href {\doibase 10.1103/PhysRevC.95.064302} {\bibfield
  {journal} {\bibinfo  {journal} {Phys. Rev. C}\ }\textbf {\bibinfo {volume}
  {95}},\ \bibinfo {pages} {064302} (\bibinfo {year} {2017})}\BibitemShut
  {NoStop}%
\bibitem [{\citenamefont {Dudouet}\ \emph {et~al.}(2017)\citenamefont
  {Dudouet}, \citenamefont {Lemasson}, \citenamefont {Duch\^ene}, \citenamefont
  {Rejmund}, \citenamefont {Cl\'ement}, \citenamefont {Michelagnoli},
  \citenamefont {Didierjean}, \citenamefont {Korichi}, \citenamefont {Maquart},
  \citenamefont {Stezowski}, \citenamefont {Lizarazo}, \citenamefont
  {P\'erez-Vidal}, \citenamefont {Andreoiu}, \citenamefont {de~Angelis},
  \citenamefont {Astier}, \citenamefont {Delafosse}, \citenamefont {Deloncle},
  \citenamefont {Dombradi}, \citenamefont {de~France}, \citenamefont {Gadea},
  \citenamefont {Gottardo}, \citenamefont {Jacquot}, \citenamefont {Jones},
  \citenamefont {Konstantinopoulos}, \citenamefont {Kuti}, \citenamefont
  {Le~Blanc}, \citenamefont {Lenzi}, \citenamefont {Li}, \citenamefont
  {Lozeva}, \citenamefont {Million}, \citenamefont {Napoli}, \citenamefont
  {Navin}, \citenamefont {Petrache}, \citenamefont {Pietralla}, \citenamefont
  {Ralet}, \citenamefont {Ramdhane}, \citenamefont {Redon}, \citenamefont
  {Schmitt}, \citenamefont {Sohler}, \citenamefont {Verney}, \citenamefont
  {Barrientos}, \citenamefont {Birkenbach}, \citenamefont {Burrows},
  \citenamefont {Charles}, \citenamefont {Collado}, \citenamefont {Cullen},
  \citenamefont {D\'esesquelles}, \citenamefont {Domingo~Pardo}, \citenamefont
  {Gonz\'alez}, \citenamefont {Harkness-Brennan}, \citenamefont {Hess},
  \citenamefont {Judson}, \citenamefont {Karolak}, \citenamefont {Korten},
  \citenamefont {Labiche}, \citenamefont {Ljungvall}, \citenamefont
  {Menegazzo}, \citenamefont {Mengoni}, \citenamefont {Pullia}, \citenamefont
  {Recchia}, \citenamefont {Reiter}, \citenamefont {Salsac}, \citenamefont
  {Sanchis}, \citenamefont {Theisen}, \citenamefont {Valiente-Dob\'on},\ and\
  \citenamefont {Zieli\ifmmode~\acute{n}\else \'{n}\fi{}ska}}]{dudouet2017}%
  \BibitemOpen
  \bibfield  {author} {\bibinfo {author} {\bibfnamefont {J.}~\bibnamefont
  {Dudouet}}, \bibinfo {author} {\bibfnamefont {A.}~\bibnamefont {Lemasson}},
  \bibinfo {author} {\bibfnamefont {G.}~\bibnamefont {Duch\^ene}}, \bibinfo
  {author} {\bibfnamefont {M.}~\bibnamefont {Rejmund}}, \bibinfo {author}
  {\bibfnamefont {E.}~\bibnamefont {Cl\'ement}}, \bibinfo {author}
  {\bibfnamefont {C.}~\bibnamefont {Michelagnoli}}, \bibinfo {author}
  {\bibfnamefont {F.}~\bibnamefont {Didierjean}}, \bibinfo {author}
  {\bibfnamefont {A.}~\bibnamefont {Korichi}}, \bibinfo {author} {\bibfnamefont
  {G.}~\bibnamefont {Maquart}}, \bibinfo {author} {\bibfnamefont
  {O.}~\bibnamefont {Stezowski}}, \bibinfo {author} {\bibfnamefont
  {C.}~\bibnamefont {Lizarazo}}, \bibinfo {author} {\bibfnamefont {R.~M.}\
  \bibnamefont {P\'erez-Vidal}}, \bibinfo {author} {\bibfnamefont
  {C.}~\bibnamefont {Andreoiu}}, \bibinfo {author} {\bibfnamefont
  {G.}~\bibnamefont {de~Angelis}}, \bibinfo {author} {\bibfnamefont
  {A.}~\bibnamefont {Astier}}, \bibinfo {author} {\bibfnamefont
  {C.}~\bibnamefont {Delafosse}}, \bibinfo {author} {\bibfnamefont
  {I.}~\bibnamefont {Deloncle}}, \bibinfo {author} {\bibfnamefont
  {Z.}~\bibnamefont {Dombradi}}, \bibinfo {author} {\bibfnamefont
  {G.}~\bibnamefont {de~France}}, \bibinfo {author} {\bibfnamefont
  {A.}~\bibnamefont {Gadea}}, \bibinfo {author} {\bibfnamefont
  {A.}~\bibnamefont {Gottardo}}, \bibinfo {author} {\bibfnamefont
  {B.}~\bibnamefont {Jacquot}}, \bibinfo {author} {\bibfnamefont
  {P.}~\bibnamefont {Jones}}, \bibinfo {author} {\bibfnamefont
  {T.}~\bibnamefont {Konstantinopoulos}}, \bibinfo {author} {\bibfnamefont
  {I.}~\bibnamefont {Kuti}}, \bibinfo {author} {\bibfnamefont {F.}~\bibnamefont
  {Le~Blanc}}, \bibinfo {author} {\bibfnamefont {S.~M.}\ \bibnamefont {Lenzi}},
  \bibinfo {author} {\bibfnamefont {G.}~\bibnamefont {Li}}, \bibinfo {author}
  {\bibfnamefont {R.}~\bibnamefont {Lozeva}}, \bibinfo {author} {\bibfnamefont
  {B.}~\bibnamefont {Million}}, \bibinfo {author} {\bibfnamefont {D.~R.}\
  \bibnamefont {Napoli}}, \bibinfo {author} {\bibfnamefont {A.}~\bibnamefont
  {Navin}}, \bibinfo {author} {\bibfnamefont {C.~M.}\ \bibnamefont {Petrache}},
  \bibinfo {author} {\bibfnamefont {N.}~\bibnamefont {Pietralla}}, \bibinfo
  {author} {\bibfnamefont {D.}~\bibnamefont {Ralet}}, \bibinfo {author}
  {\bibfnamefont {M.}~\bibnamefont {Ramdhane}}, \bibinfo {author}
  {\bibfnamefont {N.}~\bibnamefont {Redon}}, \bibinfo {author} {\bibfnamefont
  {C.}~\bibnamefont {Schmitt}}, \bibinfo {author} {\bibfnamefont
  {D.}~\bibnamefont {Sohler}}, \bibinfo {author} {\bibfnamefont
  {D.}~\bibnamefont {Verney}}, \bibinfo {author} {\bibfnamefont
  {D.}~\bibnamefont {Barrientos}}, \bibinfo {author} {\bibfnamefont
  {B.}~\bibnamefont {Birkenbach}}, \bibinfo {author} {\bibfnamefont
  {I.}~\bibnamefont {Burrows}}, \bibinfo {author} {\bibfnamefont
  {L.}~\bibnamefont {Charles}}, \bibinfo {author} {\bibfnamefont
  {J.}~\bibnamefont {Collado}}, \bibinfo {author} {\bibfnamefont {D.~M.}\
  \bibnamefont {Cullen}}, \bibinfo {author} {\bibfnamefont {P.}~\bibnamefont
  {D\'esesquelles}}, \bibinfo {author} {\bibfnamefont {C.}~\bibnamefont
  {Domingo~Pardo}}, \bibinfo {author} {\bibfnamefont {V.}~\bibnamefont
  {Gonz\'alez}}, \bibinfo {author} {\bibfnamefont {L.}~\bibnamefont
  {Harkness-Brennan}}, \bibinfo {author} {\bibfnamefont {H.}~\bibnamefont
  {Hess}}, \bibinfo {author} {\bibfnamefont {D.~S.}\ \bibnamefont {Judson}},
  \bibinfo {author} {\bibfnamefont {M.}~\bibnamefont {Karolak}}, \bibinfo
  {author} {\bibfnamefont {W.}~\bibnamefont {Korten}}, \bibinfo {author}
  {\bibfnamefont {M.}~\bibnamefont {Labiche}}, \bibinfo {author} {\bibfnamefont
  {J.}~\bibnamefont {Ljungvall}}, \bibinfo {author} {\bibfnamefont
  {R.}~\bibnamefont {Menegazzo}}, \bibinfo {author} {\bibfnamefont
  {D.}~\bibnamefont {Mengoni}}, \bibinfo {author} {\bibfnamefont
  {A.}~\bibnamefont {Pullia}}, \bibinfo {author} {\bibfnamefont
  {F.}~\bibnamefont {Recchia}}, \bibinfo {author} {\bibfnamefont
  {P.}~\bibnamefont {Reiter}}, \bibinfo {author} {\bibfnamefont {M.~D.}\
  \bibnamefont {Salsac}}, \bibinfo {author} {\bibfnamefont {E.}~\bibnamefont
  {Sanchis}}, \bibinfo {author} {\bibfnamefont {C.}~\bibnamefont {Theisen}},
  \bibinfo {author} {\bibfnamefont {J.~J.}\ \bibnamefont {Valiente-Dob\'on}}, \
  and\ \bibinfo {author} {\bibfnamefont {M.}~\bibnamefont
  {Zieli\ifmmode~\acute{n}\else \'{n}\fi{}ska}},\ }\href {\doibase
  10.1103/PhysRevLett.118.162501} {\bibfield  {journal} {\bibinfo  {journal}
  {Phys. Rev. Lett.}\ }\textbf {\bibinfo {volume} {118}},\ \bibinfo {pages}
  {162501} (\bibinfo {year} {2017})}\BibitemShut {NoStop}%
\bibitem [{\citenamefont {Flavigny}\ \emph {et~al.}(2017)\citenamefont
  {Flavigny}, \citenamefont {Doornenbal}, \citenamefont {Obertelli},
  \citenamefont {Delaroche}, \citenamefont {Girod}, \citenamefont {Libert},
  \citenamefont {Rodriguez}, \citenamefont {Authelet}, \citenamefont {Baba},
  \citenamefont {Calvet}, \citenamefont {Ch\^ateau}, \citenamefont {Chen},
  \citenamefont {Corsi}, \citenamefont {Delbart}, \citenamefont {Gheller},
  \citenamefont {Giganon}, \citenamefont {Gillibert}, \citenamefont {Lapoux},
  \citenamefont {Motobayashi}, \citenamefont {Niikura}, \citenamefont {Paul},
  \citenamefont {Rouss\'e}, \citenamefont {Sakurai}, \citenamefont
  {Santamaria}, \citenamefont {Steppenbeck}, \citenamefont {Taniuchi},
  \citenamefont {Uesaka}, \citenamefont {Ando}, \citenamefont {Arici},
  \citenamefont {Blazhev}, \citenamefont {Browne}, \citenamefont {Bruce},
  \citenamefont {Carroll}, \citenamefont {Chung}, \citenamefont {Cort\'es},
  \citenamefont {Dewald}, \citenamefont {Ding}, \citenamefont {Franchoo},
  \citenamefont {G\'orska}, \citenamefont {Gottardo}, \citenamefont
  {Jungclaus}, \citenamefont {Lee}, \citenamefont {Lettmann}, \citenamefont
  {Linh}, \citenamefont {Liu}, \citenamefont {Liu}, \citenamefont {Lizarazo},
  \citenamefont {Momiyama}, \citenamefont {Moschner}, \citenamefont {Nagamine},
  \citenamefont {Nakatsuka}, \citenamefont {Nita}, \citenamefont {Nobs},
  \citenamefont {Olivier}, \citenamefont {Orlandi}, \citenamefont {Patel},
  \citenamefont {Podoly\'ak}, \citenamefont {Rudigier}, \citenamefont {Saito},
  \citenamefont {Shand}, \citenamefont {S\"oderstr\"om}, \citenamefont
  {Stefan}, \citenamefont {Vaquero}, \citenamefont {Werner}, \citenamefont
  {Wimmer},\ and\ \citenamefont {Xu}}]{flavigny2017}%
  \BibitemOpen
  \bibfield  {author} {\bibinfo {author} {\bibfnamefont {F.}~\bibnamefont
  {Flavigny}}, \bibinfo {author} {\bibfnamefont {P.}~\bibnamefont
  {Doornenbal}}, \bibinfo {author} {\bibfnamefont {A.}~\bibnamefont
  {Obertelli}}, \bibinfo {author} {\bibfnamefont {J.-P.}\ \bibnamefont
  {Delaroche}}, \bibinfo {author} {\bibfnamefont {M.}~\bibnamefont {Girod}},
  \bibinfo {author} {\bibfnamefont {J.}~\bibnamefont {Libert}}, \bibinfo
  {author} {\bibfnamefont {T.~R.}\ \bibnamefont {Rodriguez}}, \bibinfo {author}
  {\bibfnamefont {G.}~\bibnamefont {Authelet}}, \bibinfo {author}
  {\bibfnamefont {H.}~\bibnamefont {Baba}}, \bibinfo {author} {\bibfnamefont
  {D.}~\bibnamefont {Calvet}}, \bibinfo {author} {\bibfnamefont
  {F.}~\bibnamefont {Ch\^ateau}}, \bibinfo {author} {\bibfnamefont
  {S.}~\bibnamefont {Chen}}, \bibinfo {author} {\bibfnamefont {A.}~\bibnamefont
  {Corsi}}, \bibinfo {author} {\bibfnamefont {A.}~\bibnamefont {Delbart}},
  \bibinfo {author} {\bibfnamefont {J.-M.}\ \bibnamefont {Gheller}}, \bibinfo
  {author} {\bibfnamefont {A.}~\bibnamefont {Giganon}}, \bibinfo {author}
  {\bibfnamefont {A.}~\bibnamefont {Gillibert}}, \bibinfo {author}
  {\bibfnamefont {V.}~\bibnamefont {Lapoux}}, \bibinfo {author} {\bibfnamefont
  {T.}~\bibnamefont {Motobayashi}}, \bibinfo {author} {\bibfnamefont
  {M.}~\bibnamefont {Niikura}}, \bibinfo {author} {\bibfnamefont
  {N.}~\bibnamefont {Paul}}, \bibinfo {author} {\bibfnamefont {J.-Y.}\
  \bibnamefont {Rouss\'e}}, \bibinfo {author} {\bibfnamefont {H.}~\bibnamefont
  {Sakurai}}, \bibinfo {author} {\bibfnamefont {C.}~\bibnamefont {Santamaria}},
  \bibinfo {author} {\bibfnamefont {D.}~\bibnamefont {Steppenbeck}}, \bibinfo
  {author} {\bibfnamefont {R.}~\bibnamefont {Taniuchi}}, \bibinfo {author}
  {\bibfnamefont {T.}~\bibnamefont {Uesaka}}, \bibinfo {author} {\bibfnamefont
  {T.}~\bibnamefont {Ando}}, \bibinfo {author} {\bibfnamefont {T.}~\bibnamefont
  {Arici}}, \bibinfo {author} {\bibfnamefont {A.}~\bibnamefont {Blazhev}},
  \bibinfo {author} {\bibfnamefont {F.}~\bibnamefont {Browne}}, \bibinfo
  {author} {\bibfnamefont {A.}~\bibnamefont {Bruce}}, \bibinfo {author}
  {\bibfnamefont {R.}~\bibnamefont {Carroll}}, \bibinfo {author} {\bibfnamefont
  {L.~X.}\ \bibnamefont {Chung}}, \bibinfo {author} {\bibfnamefont {M.~L.}\
  \bibnamefont {Cort\'es}}, \bibinfo {author} {\bibfnamefont {M.}~\bibnamefont
  {Dewald}}, \bibinfo {author} {\bibfnamefont {B.}~\bibnamefont {Ding}},
  \bibinfo {author} {\bibfnamefont {S.}~\bibnamefont {Franchoo}}, \bibinfo
  {author} {\bibfnamefont {M.}~\bibnamefont {G\'orska}}, \bibinfo {author}
  {\bibfnamefont {A.}~\bibnamefont {Gottardo}}, \bibinfo {author}
  {\bibfnamefont {A.}~\bibnamefont {Jungclaus}}, \bibinfo {author}
  {\bibfnamefont {J.}~\bibnamefont {Lee}}, \bibinfo {author} {\bibfnamefont
  {M.}~\bibnamefont {Lettmann}}, \bibinfo {author} {\bibfnamefont {B.~D.}\
  \bibnamefont {Linh}}, \bibinfo {author} {\bibfnamefont {J.}~\bibnamefont
  {Liu}}, \bibinfo {author} {\bibfnamefont {Z.}~\bibnamefont {Liu}}, \bibinfo
  {author} {\bibfnamefont {C.}~\bibnamefont {Lizarazo}}, \bibinfo {author}
  {\bibfnamefont {S.}~\bibnamefont {Momiyama}}, \bibinfo {author}
  {\bibfnamefont {K.}~\bibnamefont {Moschner}}, \bibinfo {author}
  {\bibfnamefont {S.}~\bibnamefont {Nagamine}}, \bibinfo {author}
  {\bibfnamefont {N.}~\bibnamefont {Nakatsuka}}, \bibinfo {author}
  {\bibfnamefont {C.}~\bibnamefont {Nita}}, \bibinfo {author} {\bibfnamefont
  {C.~R.}\ \bibnamefont {Nobs}}, \bibinfo {author} {\bibfnamefont
  {L.}~\bibnamefont {Olivier}}, \bibinfo {author} {\bibfnamefont
  {R.}~\bibnamefont {Orlandi}}, \bibinfo {author} {\bibfnamefont
  {Z.}~\bibnamefont {Patel}}, \bibinfo {author} {\bibfnamefont
  {Z.}~\bibnamefont {Podoly\'ak}}, \bibinfo {author} {\bibfnamefont
  {M.}~\bibnamefont {Rudigier}}, \bibinfo {author} {\bibfnamefont
  {T.}~\bibnamefont {Saito}}, \bibinfo {author} {\bibfnamefont
  {C.}~\bibnamefont {Shand}}, \bibinfo {author} {\bibfnamefont {P.~A.}\
  \bibnamefont {S\"oderstr\"om}}, \bibinfo {author} {\bibfnamefont
  {I.}~\bibnamefont {Stefan}}, \bibinfo {author} {\bibfnamefont
  {V.}~\bibnamefont {Vaquero}}, \bibinfo {author} {\bibfnamefont
  {V.}~\bibnamefont {Werner}}, \bibinfo {author} {\bibfnamefont
  {K.}~\bibnamefont {Wimmer}}, \ and\ \bibinfo {author} {\bibfnamefont
  {Z.}~\bibnamefont {Xu}},\ }\href {\doibase 10.1103/PhysRevLett.118.242501}
  {\bibfield  {journal} {\bibinfo  {journal} {Phys. Rev. Lett.}\ }\textbf
  {\bibinfo {volume} {118}},\ \bibinfo {pages} {242501} (\bibinfo {year}
  {2017})}\BibitemShut {NoStop}%
\bibitem [{\citenamefont {Togashi}\ \emph {et~al.}(2016)\citenamefont
  {Togashi}, \citenamefont {Tsunoda}, \citenamefont {Otsuka},\ and\
  \citenamefont {Shimizu}}]{togashi2016}%
  \BibitemOpen
  \bibfield  {author} {\bibinfo {author} {\bibfnamefont {T.}~\bibnamefont
  {Togashi}}, \bibinfo {author} {\bibfnamefont {Y.}~\bibnamefont {Tsunoda}},
  \bibinfo {author} {\bibfnamefont {T.}~\bibnamefont {Otsuka}}, \ and\ \bibinfo
  {author} {\bibfnamefont {N.}~\bibnamefont {Shimizu}},\ }\href {\doibase
  10.1103/PhysRevLett.117.172502} {\bibfield  {journal} {\bibinfo  {journal}
  {Phys. Rev. Lett.}\ }\textbf {\bibinfo {volume} {117}},\ \bibinfo {pages}
  {172502} (\bibinfo {year} {2016})}\BibitemShut {NoStop}%
\bibitem [{\citenamefont {Kremer}\ \emph {et~al.}(2016)\citenamefont {Kremer},
  \citenamefont {Aslanidou}, \citenamefont {Bassauer}, \citenamefont {Hilcker},
  \citenamefont {Krugmann}, \citenamefont {von Neumann-Cosel}, \citenamefont
  {Otsuka}, \citenamefont {Pietralla}, \citenamefont {Ponomarev}, \citenamefont
  {Shimizu}, \citenamefont {Singer}, \citenamefont {Steinhilber}, \citenamefont
  {Togashi}, \citenamefont {Tsunoda}, \citenamefont {Werner},\ and\
  \citenamefont {Zweidinger}}]{kremer2016}%
  \BibitemOpen
  \bibfield  {author} {\bibinfo {author} {\bibfnamefont {C.}~\bibnamefont
  {Kremer}}, \bibinfo {author} {\bibfnamefont {S.}~\bibnamefont {Aslanidou}},
  \bibinfo {author} {\bibfnamefont {S.}~\bibnamefont {Bassauer}}, \bibinfo
  {author} {\bibfnamefont {M.}~\bibnamefont {Hilcker}}, \bibinfo {author}
  {\bibfnamefont {A.}~\bibnamefont {Krugmann}}, \bibinfo {author}
  {\bibfnamefont {P.}~\bibnamefont {von Neumann-Cosel}}, \bibinfo {author}
  {\bibfnamefont {T.}~\bibnamefont {Otsuka}}, \bibinfo {author} {\bibfnamefont
  {N.}~\bibnamefont {Pietralla}}, \bibinfo {author} {\bibfnamefont {V.~Y.}\
  \bibnamefont {Ponomarev}}, \bibinfo {author} {\bibfnamefont {N.}~\bibnamefont
  {Shimizu}}, \bibinfo {author} {\bibfnamefont {M.}~\bibnamefont {Singer}},
  \bibinfo {author} {\bibfnamefont {G.}~\bibnamefont {Steinhilber}}, \bibinfo
  {author} {\bibfnamefont {T.}~\bibnamefont {Togashi}}, \bibinfo {author}
  {\bibfnamefont {Y.}~\bibnamefont {Tsunoda}}, \bibinfo {author} {\bibfnamefont
  {V.}~\bibnamefont {Werner}}, \ and\ \bibinfo {author} {\bibfnamefont
  {M.}~\bibnamefont {Zweidinger}},\ }\href {\doibase
  10.1103/PhysRevLett.117.172503} {\bibfield  {journal} {\bibinfo  {journal}
  {Phys. Rev. Lett.}\ }\textbf {\bibinfo {volume} {117}},\ \bibinfo {pages}
  {172503} (\bibinfo {year} {2016})}\BibitemShut {NoStop}%
\bibitem [{\citenamefont {Cl\'ement}\ \emph {et~al.}(2016)\citenamefont
  {Cl\'ement}, \citenamefont {Zieli\ifmmode~\acute{n}\else \'{n}\fi{}ska},
  \citenamefont {P\'eru}, \citenamefont {Goutte}, \citenamefont {Hilaire},
  \citenamefont {G\"orgen}, \citenamefont {Korten}, \citenamefont {Doherty},
  \citenamefont {Bastin}, \citenamefont {Bauer}, \citenamefont {Blazhev},
  \citenamefont {Bree}, \citenamefont {Bruyneel}, \citenamefont {Butler},
  \citenamefont {Butterworth}, \citenamefont {Cederk\"all}, \citenamefont
  {Delahaye}, \citenamefont {Dijon}, \citenamefont {Ekstr\"om}, \citenamefont
  {Fitzpatrick}, \citenamefont {Fransen}, \citenamefont {Georgiev},
  \citenamefont {Gernh\"auser}, \citenamefont {Hess}, \citenamefont {Iwanicki},
  \citenamefont {Jenkins}, \citenamefont {Larsen}, \citenamefont {Ljungvall},
  \citenamefont {Lutter}, \citenamefont {Marley}, \citenamefont {Moschner},
  \citenamefont {Napiorkowski}, \citenamefont {Pakarinen}, \citenamefont
  {Petts}, \citenamefont {Reiter}, \citenamefont {Renstr\o{}m}, \citenamefont
  {Seidlitz}, \citenamefont {Siebeck}, \citenamefont {Siem}, \citenamefont
  {Sotty}, \citenamefont {Srebrny}, \citenamefont {Stefanescu}, \citenamefont
  {Tveten}, \citenamefont {Van~de Walle}, \citenamefont {Vermeulen},
  \citenamefont {Voulot}, \citenamefont {Warr}, \citenamefont {Wenander},
  \citenamefont {Wiens}, \citenamefont {De~Witte},\ and\ \citenamefont
  {Wrzosek-Lipska}}]{clement2017}%
  \BibitemOpen
  \bibfield  {author} {\bibinfo {author} {\bibfnamefont {E.}~\bibnamefont
  {Cl\'ement}}, \bibinfo {author} {\bibfnamefont {M.}~\bibnamefont
  {Zieli\ifmmode~\acute{n}\else \'{n}\fi{}ska}}, \bibinfo {author}
  {\bibfnamefont {S.}~\bibnamefont {P\'eru}}, \bibinfo {author} {\bibfnamefont
  {H.}~\bibnamefont {Goutte}}, \bibinfo {author} {\bibfnamefont
  {S.}~\bibnamefont {Hilaire}}, \bibinfo {author} {\bibfnamefont
  {A.}~\bibnamefont {G\"orgen}}, \bibinfo {author} {\bibfnamefont
  {W.}~\bibnamefont {Korten}}, \bibinfo {author} {\bibfnamefont {D.~T.}\
  \bibnamefont {Doherty}}, \bibinfo {author} {\bibfnamefont {B.}~\bibnamefont
  {Bastin}}, \bibinfo {author} {\bibfnamefont {C.}~\bibnamefont {Bauer}},
  \bibinfo {author} {\bibfnamefont {A.}~\bibnamefont {Blazhev}}, \bibinfo
  {author} {\bibfnamefont {N.}~\bibnamefont {Bree}}, \bibinfo {author}
  {\bibfnamefont {B.}~\bibnamefont {Bruyneel}}, \bibinfo {author}
  {\bibfnamefont {P.~A.}\ \bibnamefont {Butler}}, \bibinfo {author}
  {\bibfnamefont {J.}~\bibnamefont {Butterworth}}, \bibinfo {author}
  {\bibfnamefont {J.}~\bibnamefont {Cederk\"all}}, \bibinfo {author}
  {\bibfnamefont {P.}~\bibnamefont {Delahaye}}, \bibinfo {author}
  {\bibfnamefont {A.}~\bibnamefont {Dijon}}, \bibinfo {author} {\bibfnamefont
  {A.}~\bibnamefont {Ekstr\"om}}, \bibinfo {author} {\bibfnamefont
  {C.}~\bibnamefont {Fitzpatrick}}, \bibinfo {author} {\bibfnamefont
  {C.}~\bibnamefont {Fransen}}, \bibinfo {author} {\bibfnamefont
  {G.}~\bibnamefont {Georgiev}}, \bibinfo {author} {\bibfnamefont
  {R.}~\bibnamefont {Gernh\"auser}}, \bibinfo {author} {\bibfnamefont
  {H.}~\bibnamefont {Hess}}, \bibinfo {author} {\bibfnamefont {J.}~\bibnamefont
  {Iwanicki}}, \bibinfo {author} {\bibfnamefont {D.~G.}\ \bibnamefont
  {Jenkins}}, \bibinfo {author} {\bibfnamefont {A.~C.}\ \bibnamefont {Larsen}},
  \bibinfo {author} {\bibfnamefont {J.}~\bibnamefont {Ljungvall}}, \bibinfo
  {author} {\bibfnamefont {R.}~\bibnamefont {Lutter}}, \bibinfo {author}
  {\bibfnamefont {P.}~\bibnamefont {Marley}}, \bibinfo {author} {\bibfnamefont
  {K.}~\bibnamefont {Moschner}}, \bibinfo {author} {\bibfnamefont {P.~J.}\
  \bibnamefont {Napiorkowski}}, \bibinfo {author} {\bibfnamefont
  {J.}~\bibnamefont {Pakarinen}}, \bibinfo {author} {\bibfnamefont
  {A.}~\bibnamefont {Petts}}, \bibinfo {author} {\bibfnamefont
  {P.}~\bibnamefont {Reiter}}, \bibinfo {author} {\bibfnamefont
  {T.}~\bibnamefont {Renstr\o{}m}}, \bibinfo {author} {\bibfnamefont
  {M.}~\bibnamefont {Seidlitz}}, \bibinfo {author} {\bibfnamefont
  {B.}~\bibnamefont {Siebeck}}, \bibinfo {author} {\bibfnamefont
  {S.}~\bibnamefont {Siem}}, \bibinfo {author} {\bibfnamefont {C.}~\bibnamefont
  {Sotty}}, \bibinfo {author} {\bibfnamefont {J.}~\bibnamefont {Srebrny}},
  \bibinfo {author} {\bibfnamefont {I.}~\bibnamefont {Stefanescu}}, \bibinfo
  {author} {\bibfnamefont {G.~M.}\ \bibnamefont {Tveten}}, \bibinfo {author}
  {\bibfnamefont {J.}~\bibnamefont {Van~de Walle}}, \bibinfo {author}
  {\bibfnamefont {M.}~\bibnamefont {Vermeulen}}, \bibinfo {author}
  {\bibfnamefont {D.}~\bibnamefont {Voulot}}, \bibinfo {author} {\bibfnamefont
  {N.}~\bibnamefont {Warr}}, \bibinfo {author} {\bibfnamefont {F.}~\bibnamefont
  {Wenander}}, \bibinfo {author} {\bibfnamefont {A.}~\bibnamefont {Wiens}},
  \bibinfo {author} {\bibfnamefont {H.}~\bibnamefont {De~Witte}}, \ and\
  \bibinfo {author} {\bibfnamefont {K.}~\bibnamefont {Wrzosek-Lipska}},\ }\href
  {\doibase 10.1103/PhysRevC.94.054326} {\bibfield  {journal} {\bibinfo
  {journal} {Phys. Rev. C}\ }\textbf {\bibinfo {volume} {94}},\ \bibinfo
  {pages} {054326} (\bibinfo {year} {2016})}\BibitemShut {NoStop}%
\bibitem [{\citenamefont {Caurier}\ \emph {et~al.}(2005)\citenamefont
  {Caurier}, \citenamefont {Mart\'inez-Pinedo}, \citenamefont {Nowacki},
  \citenamefont {Poves},\ and\ \citenamefont {Zuker}}]{caurier2005}%
  \BibitemOpen
  \bibfield  {author} {\bibinfo {author} {\bibfnamefont {E.}~\bibnamefont
  {Caurier}}, \bibinfo {author} {\bibfnamefont {G.}~\bibnamefont
  {Mart\'inez-Pinedo}}, \bibinfo {author} {\bibfnamefont {F.}~\bibnamefont
  {Nowacki}}, \bibinfo {author} {\bibfnamefont {A.}~\bibnamefont {Poves}}, \
  and\ \bibinfo {author} {\bibfnamefont {A.~P.}\ \bibnamefont {Zuker}},\
  }\href@noop {} {\bibfield  {journal} {\bibinfo  {journal} {Rev. Mod. Phys.}\
  }\textbf {\bibinfo {volume} {77}},\ \bibinfo {pages} {427} (\bibinfo {year}
  {2005})}\BibitemShut {NoStop}%
\bibitem [{\citenamefont {Bender}\ \emph {et~al.}(2003)\citenamefont {Bender},
  \citenamefont {Heenen},\ and\ \citenamefont {Reinhard}}]{bender2003}%
  \BibitemOpen
  \bibfield  {author} {\bibinfo {author} {\bibfnamefont {M.}~\bibnamefont
  {Bender}}, \bibinfo {author} {\bibfnamefont {P.-H.}\ \bibnamefont {Heenen}},
  \ and\ \bibinfo {author} {\bibfnamefont {P.-G.}\ \bibnamefont {Reinhard}},\
  }\href {\doibase 10.1103/RevModPhys.75.121} {\bibfield  {journal} {\bibinfo
  {journal} {Rev. Mod. Phys.}\ }\textbf {\bibinfo {volume} {75}},\ \bibinfo
  {pages} {121} (\bibinfo {year} {2003})}\BibitemShut {NoStop}%
\bibitem [{\citenamefont {Vretenar}\ \emph {et~al.}(2005)\citenamefont
  {Vretenar}, \citenamefont {Afanasjev}, \citenamefont {Lalazissis},\ and\
  \citenamefont {Ring}}]{vretenar2005}%
  \BibitemOpen
  \bibfield  {author} {\bibinfo {author} {\bibfnamefont {D.}~\bibnamefont
  {Vretenar}}, \bibinfo {author} {\bibfnamefont {A.~V.}\ \bibnamefont
  {Afanasjev}}, \bibinfo {author} {\bibfnamefont {G.}~\bibnamefont
  {Lalazissis}}, \ and\ \bibinfo {author} {\bibfnamefont {P.}~\bibnamefont
  {Ring}},\ }\href {\doibase 10.1016/j.physrep.2004.10.001} {\bibfield
  {journal} {\bibinfo  {journal} {Phys. Rep.}\ }\textbf {\bibinfo {volume}
  {409}},\ \bibinfo {pages} {101} (\bibinfo {year} {2005})}\BibitemShut
  {NoStop}%
\bibitem [{\citenamefont {Nik\ifmmode \check{s}\else
  \v{s}\fi{}i\ifmmode~\acute{c}\else \'{c}\fi{}}\ \emph
  {et~al.}(2011)\citenamefont {Nik\ifmmode \check{s}\else
  \v{s}\fi{}i\ifmmode~\acute{c}\else \'{c}\fi{}}, \citenamefont {Vretenar},\
  and\ \citenamefont {Ring}}]{niksic2011}%
  \BibitemOpen
  \bibfield  {author} {\bibinfo {author} {\bibfnamefont {T.}~\bibnamefont
  {Nik\ifmmode \check{s}\else \v{s}\fi{}i\ifmmode~\acute{c}\else \'{c}\fi{}}},
  \bibinfo {author} {\bibfnamefont {D.}~\bibnamefont {Vretenar}}, \ and\
  \bibinfo {author} {\bibfnamefont {P.}~\bibnamefont {Ring}},\ }\href {\doibase
  10.1016/j.ppnp.2011.01.055} {\bibfield  {journal} {\bibinfo  {journal} {Prog.
  Part. Nucl. Phys.}\ }\textbf {\bibinfo {volume} {66}},\ \bibinfo {pages}
  {519} (\bibinfo {year} {2011})}\BibitemShut {NoStop}%
\bibitem [{\citenamefont {Rodr\'iguez-Guzm\'an}\ \emph
  {et~al.}(2002)\citenamefont {Rodr\'iguez-Guzm\'an}, \citenamefont {Egido},\
  and\ \citenamefont {Robledo}}]{rayner2002}%
  \BibitemOpen
  \bibfield  {author} {\bibinfo {author} {\bibfnamefont {R.}~\bibnamefont
  {Rodr\'iguez-Guzm\'an}}, \bibinfo {author} {\bibfnamefont {J.~L.}\
  \bibnamefont {Egido}}, \ and\ \bibinfo {author} {\bibfnamefont {L.~M.}\
  \bibnamefont {Robledo}},\ }\href@noop {} {\bibfield  {journal} {\bibinfo
  {journal} {Nucl. Phys. A}\ }\textbf {\bibinfo {volume} {709}},\ \bibinfo
  {pages} {201 } (\bibinfo {year} {2002})}\BibitemShut {NoStop}%
\bibitem [{\citenamefont {{J. -P. Delaroche {\it et
  al.}}}(2010)}]{delaroche2010}%
  \BibitemOpen
  \bibfield  {author} {\bibinfo {author} {\bibnamefont {{J. -P. Delaroche {\it
  et al.}}}},\ }\href {\doibase 10.1103/PhysRevC.81.014303} {\bibfield
  {journal} {\bibinfo  {journal} {Phys. Rev. C}\ }\textbf {\bibinfo {volume}
  {81}},\ \bibinfo {pages} {014303} (\bibinfo {year} {2010})}\BibitemShut
  {NoStop}%
\bibitem [{\citenamefont {Nomura}\ \emph {et~al.}(2008)\citenamefont {Nomura},
  \citenamefont {Shimizu},\ and\ \citenamefont {Otsuka}}]{nomura2008}%
  \BibitemOpen
  \bibfield  {author} {\bibinfo {author} {\bibfnamefont {K.}~\bibnamefont
  {Nomura}}, \bibinfo {author} {\bibfnamefont {N.}~\bibnamefont {Shimizu}}, \
  and\ \bibinfo {author} {\bibfnamefont {T.}~\bibnamefont {Otsuka}},\ }\href
  {\doibase 10.1103/PhysRevLett.101.142501} {\bibfield  {journal} {\bibinfo
  {journal} {Phys. Rev. Lett.}\ }\textbf {\bibinfo {volume} {101}},\ \bibinfo
  {pages} {142501} (\bibinfo {year} {2008})}\BibitemShut {NoStop}%
\bibitem [{\citenamefont {Ring}\ and\ \citenamefont {Schuck}(1980)}]{RS}%
  \BibitemOpen
  \bibfield  {author} {\bibinfo {author} {\bibfnamefont {P.}~\bibnamefont
  {Ring}}\ and\ \bibinfo {author} {\bibfnamefont {P.}~\bibnamefont {Schuck}},\
  }\href@noop {} {\emph {\bibinfo {title} {The nuclear many-body problem}}}\
  (\bibinfo  {publisher} {Berlin: Springer-Verlag},\ \bibinfo {year}
  {1980})\BibitemShut {NoStop}%
\bibitem [{\citenamefont {Nomura}\ \emph {et~al.}(2016)\citenamefont {Nomura},
  \citenamefont {Rodr\'{\i}guez-Guzm\'an},\ and\ \citenamefont
  {Robledo}}]{nomura2016zr}%
  \BibitemOpen
  \bibfield  {author} {\bibinfo {author} {\bibfnamefont {K.}~\bibnamefont
  {Nomura}}, \bibinfo {author} {\bibfnamefont {R.}~\bibnamefont
  {Rodr\'{\i}guez-Guzm\'an}}, \ and\ \bibinfo {author} {\bibfnamefont {L.~M.}\
  \bibnamefont {Robledo}},\ }\href {\doibase 10.1103/PhysRevC.94.044314}
  {\bibfield  {journal} {\bibinfo  {journal} {Phys. Rev. C}\ }\textbf {\bibinfo
  {volume} {94}},\ \bibinfo {pages} {044314} (\bibinfo {year}
  {2016})}\BibitemShut {NoStop}%
\bibitem [{\citenamefont {Nomura}\ \emph {et~al.}(2017)\citenamefont {Nomura},
  \citenamefont {Rodr\'{\i}guez-Guzm\'an},\ and\ \citenamefont
  {Robledo}}]{nomura2017ge}%
  \BibitemOpen
  \bibfield  {author} {\bibinfo {author} {\bibfnamefont {K.}~\bibnamefont
  {Nomura}}, \bibinfo {author} {\bibfnamefont {R.}~\bibnamefont
  {Rodr\'{\i}guez-Guzm\'an}}, \ and\ \bibinfo {author} {\bibfnamefont {L.~M.}\
  \bibnamefont {Robledo}},\ }\href {\doibase 10.1103/PhysRevC.95.064310}
  {\bibfield  {journal} {\bibinfo  {journal} {Phys. Rev. C}\ }\textbf {\bibinfo
  {volume} {95}},\ \bibinfo {pages} {064310} (\bibinfo {year}
  {2017})}\BibitemShut {NoStop}%
\bibitem [{\citenamefont {Goriely}\ \emph {et~al.}(2009)\citenamefont
  {Goriely}, \citenamefont {Hilaire}, \citenamefont {Girod},\ and\
  \citenamefont {P\'eru}}]{D1M}%
  \BibitemOpen
  \bibfield  {author} {\bibinfo {author} {\bibfnamefont {S.}~\bibnamefont
  {Goriely}}, \bibinfo {author} {\bibfnamefont {S.}~\bibnamefont {Hilaire}},
  \bibinfo {author} {\bibfnamefont {M.}~\bibnamefont {Girod}}, \ and\ \bibinfo
  {author} {\bibfnamefont {S.}~\bibnamefont {P\'eru}},\ }\href {\doibase
  10.1103/PhysRevLett.102.242501} {\bibfield  {journal} {\bibinfo  {journal}
  {Phys. Rev. Lett.}\ }\textbf {\bibinfo {volume} {102}},\ \bibinfo {pages}
  {242501} (\bibinfo {year} {2009})}\BibitemShut {NoStop}%
\bibitem [{\citenamefont {Berger}\ \emph {et~al.}(1984)\citenamefont {Berger},
  \citenamefont {Girod},\ and\ \citenamefont {Gogny}}]{D1S}%
  \BibitemOpen
  \bibfield  {author} {\bibinfo {author} {\bibfnamefont {J.~F.}\ \bibnamefont
  {Berger}}, \bibinfo {author} {\bibfnamefont {M.}~\bibnamefont {Girod}}, \
  and\ \bibinfo {author} {\bibfnamefont {D.}~\bibnamefont {Gogny}},\
  }\href@noop {} {\bibfield  {journal} {\bibinfo  {journal} {Nucl. Phys. A}\
  }\textbf {\bibinfo {volume} {428}},\ \bibinfo {pages} {23 } (\bibinfo {year}
  {1984})}\BibitemShut {NoStop}%
\bibitem [{\citenamefont {Chappert}\ \emph {et~al.}(2008)\citenamefont
  {Chappert}, \citenamefont {Girod},\ and\ \citenamefont {Hilaire}}]{D1N}%
  \BibitemOpen
  \bibfield  {author} {\bibinfo {author} {\bibfnamefont {F.}~\bibnamefont
  {Chappert}}, \bibinfo {author} {\bibfnamefont {M.}~\bibnamefont {Girod}}, \
  and\ \bibinfo {author} {\bibfnamefont {S.}~\bibnamefont {Hilaire}},\ }\href
  {\doibase https://doi.org/10.1016/j.physletb.2008.09.017} {\bibfield
  {journal} {\bibinfo  {journal} {Physics Letters B}\ }\textbf {\bibinfo
  {volume} {668}},\ \bibinfo {pages} {420 } (\bibinfo {year}
  {2008})}\BibitemShut {NoStop}%
\bibitem [{\citenamefont {Lalazissis}\ \emph {et~al.}(2005)\citenamefont
  {Lalazissis}, \citenamefont {Nik\ifmmode \check{s}\else
  \v{s}\fi{}i\ifmmode~\acute{c}\else \'{c}\fi{}}, \citenamefont {Vretenar},\
  and\ \citenamefont {Ring}}]{lalazissis2005}%
  \BibitemOpen
  \bibfield  {author} {\bibinfo {author} {\bibfnamefont {G.~A.}\ \bibnamefont
  {Lalazissis}}, \bibinfo {author} {\bibfnamefont {T.}~\bibnamefont
  {Nik\ifmmode \check{s}\else \v{s}\fi{}i\ifmmode~\acute{c}\else \'{c}\fi{}}},
  \bibinfo {author} {\bibfnamefont {D.}~\bibnamefont {Vretenar}}, \ and\
  \bibinfo {author} {\bibfnamefont {P.}~\bibnamefont {Ring}},\ }\href {\doibase
  10.1103/PhysRevC.71.024312} {\bibfield  {journal} {\bibinfo  {journal} {Phys.
  Rev. C}\ }\textbf {\bibinfo {volume} {71}},\ \bibinfo {pages} {024312}
  (\bibinfo {year} {2005})}\BibitemShut {NoStop}%
\bibitem [{\citenamefont {Nik\ifmmode \check{s}\else
  \v{s}\fi{}i\ifmmode~\acute{c}\else \'{c}\fi{}}\ \emph
  {et~al.}(2008)\citenamefont {Nik\ifmmode \check{s}\else
  \v{s}\fi{}i\ifmmode~\acute{c}\else \'{c}\fi{}}, \citenamefont {Vretenar},\
  and\ \citenamefont {Ring}}]{DDPC1}%
  \BibitemOpen
  \bibfield  {author} {\bibinfo {author} {\bibfnamefont {T.}~\bibnamefont
  {Nik\ifmmode \check{s}\else \v{s}\fi{}i\ifmmode~\acute{c}\else \'{c}\fi{}}},
  \bibinfo {author} {\bibfnamefont {D.}~\bibnamefont {Vretenar}}, \ and\
  \bibinfo {author} {\bibfnamefont {P.}~\bibnamefont {Ring}},\ }\href {\doibase
  10.1103/PhysRevC.78.034318} {\bibfield  {journal} {\bibinfo  {journal} {Phys.
  Rev. C}\ }\textbf {\bibinfo {volume} {78}},\ \bibinfo {pages} {034318}
  (\bibinfo {year} {2008})}\BibitemShut {NoStop}%
\bibitem [{\citenamefont {Rodr\'iguez-Guzm\'an}\ \emph
  {et~al.}(2010)\citenamefont {Rodr\'iguez-Guzm\'an}, \citenamefont
  {Sarriguren}, \citenamefont {Robledo},\ and\ \citenamefont
  {Garc\'ia-Ramos}}]{rayner2010pt}%
  \BibitemOpen
  \bibfield  {author} {\bibinfo {author} {\bibfnamefont {R.}~\bibnamefont
  {Rodr\'iguez-Guzm\'an}}, \bibinfo {author} {\bibfnamefont {P.}~\bibnamefont
  {Sarriguren}}, \bibinfo {author} {\bibfnamefont {L.~M.}\ \bibnamefont
  {Robledo}}, \ and\ \bibinfo {author} {\bibfnamefont {J.~E.}\ \bibnamefont
  {Garc\'ia-Ramos}},\ }\href {\doibase 10.1103/PhysRevC.81.024310} {\bibfield
  {journal} {\bibinfo  {journal} {Phys. Rev. C}\ }\textbf {\bibinfo {volume}
  {81}},\ \bibinfo {pages} {024310} (\bibinfo {year} {2010})}\BibitemShut
  {NoStop}%
\bibitem [{\citenamefont {Bohr}\ and\ \citenamefont {Mottelsson}(1975)}]{BM}%
  \BibitemOpen
  \bibfield  {author} {\bibinfo {author} {\bibfnamefont {A.}~\bibnamefont
  {Bohr}}\ and\ \bibinfo {author} {\bibfnamefont {B.~M.}\ \bibnamefont
  {Mottelsson}},\ }\href@noop {} {\emph {\bibinfo {title} {Nuclear
  Structure}}},\ Vol.~\bibinfo {volume} {2}\ (\bibinfo  {publisher} {Benjamin,
  New York, USA},\ \bibinfo {year} {1975})\ p.~\bibinfo {pages}
  {45}\BibitemShut {NoStop}%
\bibitem [{\citenamefont {Iachello}\ and\ \citenamefont {Arima}(1987)}]{IBM}%
  \BibitemOpen
  \bibfield  {author} {\bibinfo {author} {\bibfnamefont {F.}~\bibnamefont
  {Iachello}}\ and\ \bibinfo {author} {\bibfnamefont {A.}~\bibnamefont
  {Arima}},\ }\href@noop {} {\emph {\bibinfo {title} {The interacting boson
  model}}}\ (\bibinfo  {publisher} {Cambridge University Press, Cambridge},\
  \bibinfo {year} {1987})\BibitemShut {NoStop}%
\bibitem [{\citenamefont {Otsuka}\ \emph {et~al.}(1978)\citenamefont {Otsuka},
  \citenamefont {Arima},\ and\ \citenamefont {Iachello}}]{OAI}%
  \BibitemOpen
  \bibfield  {author} {\bibinfo {author} {\bibfnamefont {T.}~\bibnamefont
  {Otsuka}}, \bibinfo {author} {\bibfnamefont {A.}~\bibnamefont {Arima}}, \
  and\ \bibinfo {author} {\bibfnamefont {F.}~\bibnamefont {Iachello}},\
  }\href@noop {} {\bibfield  {journal} {\bibinfo  {journal} {Nucl. Phys. A}\
  }\textbf {\bibinfo {volume} {309}},\ \bibinfo {pages} {1} (\bibinfo {year}
  {1978})}\BibitemShut {NoStop}%
\bibitem [{\citenamefont {Kaup}\ and\ \citenamefont
  {Gelberg}(1979)}]{kaup1979}%
  \BibitemOpen
  \bibfield  {author} {\bibinfo {author} {\bibfnamefont {U.}~\bibnamefont
  {Kaup}}\ and\ \bibinfo {author} {\bibfnamefont {A.}~\bibnamefont {Gelberg}},\
  }\href {\doibase 10.1007/BF01435273} {\bibfield  {journal} {\bibinfo
  {journal} {Zeitschrift fur Physik A Atoms and Nuclei}\ }\textbf {\bibinfo
  {volume} {293}},\ \bibinfo {pages} {311} (\bibinfo {year}
  {1979})}\BibitemShut {NoStop}%
\bibitem [{\citenamefont {Duval}\ \emph {et~al.}(1983)\citenamefont {Duval},
  \citenamefont {Goutte},\ and\ \citenamefont {Vergnes}}]{duval1983}%
  \BibitemOpen
  \bibfield  {author} {\bibinfo {author} {\bibfnamefont {P.}~\bibnamefont
  {Duval}}, \bibinfo {author} {\bibfnamefont {D.}~\bibnamefont {Goutte}}, \
  and\ \bibinfo {author} {\bibfnamefont {M.}~\bibnamefont {Vergnes}},\ }\href
  {\doibase http://dx.doi.org/10.1016/0370-2693(83)91457-0} {\bibfield
  {journal} {\bibinfo  {journal} {Physics Letters B}\ }\textbf {\bibinfo
  {volume} {124}},\ \bibinfo {pages} {297 } (\bibinfo {year}
  {1983})}\BibitemShut {NoStop}%
\bibitem [{\citenamefont {Padilla-Rodal}\ \emph {et~al.}(2006)\citenamefont
  {Padilla-Rodal}, \citenamefont {Castanos}, \citenamefont {Bijker},\ and\
  \citenamefont {Galindo-Uribarri}}]{padilla2006}%
  \BibitemOpen
  \bibfield  {author} {\bibinfo {author} {\bibfnamefont {E.}~\bibnamefont
  {Padilla-Rodal}}, \bibinfo {author} {\bibfnamefont {O.}~\bibnamefont
  {Castanos}}, \bibinfo {author} {\bibfnamefont {R.}~\bibnamefont {Bijker}}, \
  and\ \bibinfo {author} {\bibfnamefont {A.}~\bibnamefont {Galindo-Uribarri}},\
  }\href@noop {} {\bibfield  {journal} {\bibinfo  {journal} {Rev. Mex. Fis. S}\
  }\textbf {\bibinfo {volume} {52}},\ \bibinfo {pages} {57} (\bibinfo {year}
  {2006})}\BibitemShut {NoStop}%
\bibitem [{\citenamefont {Duval}\ and\ \citenamefont
  {Barrett}(1981)}]{duval1981}%
  \BibitemOpen
  \bibfield  {author} {\bibinfo {author} {\bibfnamefont {P.~D.}\ \bibnamefont
  {Duval}}\ and\ \bibinfo {author} {\bibfnamefont {B.~R.}\ \bibnamefont
  {Barrett}},\ }\href@noop {} {\bibfield  {journal} {\bibinfo  {journal} {Phys.
  Lett. B}\ }\textbf {\bibinfo {volume} {100}},\ \bibinfo {pages} {223}
  (\bibinfo {year} {1981})}\BibitemShut {NoStop}%
\bibitem [{\citenamefont {Duval}\ and\ \citenamefont
  {Barrett}(1982)}]{duval1982}%
  \BibitemOpen
  \bibfield  {author} {\bibinfo {author} {\bibfnamefont {P.~D.}\ \bibnamefont
  {Duval}}\ and\ \bibinfo {author} {\bibfnamefont {B.~R.}\ \bibnamefont
  {Barrett}},\ }\href@noop {} {\bibfield  {journal} {\bibinfo  {journal} {Nucl.
  Phys. A}\ }\textbf {\bibinfo {volume} {376}},\ \bibinfo {pages} {213 }
  (\bibinfo {year} {1982})}\BibitemShut {NoStop}%
\bibitem [{\citenamefont {Van~Isacker}\ and\ \citenamefont
  {Chen}(1981)}]{vanisacker1981}%
  \BibitemOpen
  \bibfield  {author} {\bibinfo {author} {\bibfnamefont {P.}~\bibnamefont
  {Van~Isacker}}\ and\ \bibinfo {author} {\bibfnamefont {J.-Q.}\ \bibnamefont
  {Chen}},\ }\href {\doibase 10.1103/PhysRevC.24.684} {\bibfield  {journal}
  {\bibinfo  {journal} {Phys. Rev. C}\ }\textbf {\bibinfo {volume} {24}},\
  \bibinfo {pages} {684} (\bibinfo {year} {1981})}\BibitemShut {NoStop}%
\bibitem [{\citenamefont {Nomura}\ \emph
  {et~al.}(2012{\natexlab{a}})\citenamefont {Nomura}, \citenamefont {Shimizu},
  \citenamefont {Vretenar}, \citenamefont {Nik\ifmmode \check{s}\else
  \v{s}\fi{}i\ifmmode~\acute{c}\else \'{c}\fi{}},\ and\ \citenamefont
  {Otsuka}}]{nomura2012tri}%
  \BibitemOpen
  \bibfield  {author} {\bibinfo {author} {\bibfnamefont {K.}~\bibnamefont
  {Nomura}}, \bibinfo {author} {\bibfnamefont {N.}~\bibnamefont {Shimizu}},
  \bibinfo {author} {\bibfnamefont {D.}~\bibnamefont {Vretenar}}, \bibinfo
  {author} {\bibfnamefont {T.}~\bibnamefont {Nik\ifmmode \check{s}\else
  \v{s}\fi{}i\ifmmode~\acute{c}\else \'{c}\fi{}}}, \ and\ \bibinfo {author}
  {\bibfnamefont {T.}~\bibnamefont {Otsuka}},\ }\href {\doibase
  10.1103/PhysRevLett.108.132501} {\bibfield  {journal} {\bibinfo  {journal}
  {Phys. Rev. Lett.}\ }\textbf {\bibinfo {volume} {108}},\ \bibinfo {pages}
  {132501} (\bibinfo {year} {2012}{\natexlab{a}})}\BibitemShut {NoStop}%
\bibitem [{\citenamefont {Frank}\ \emph {et~al.}(2004)\citenamefont {Frank},
  \citenamefont {Van~Isacker},\ and\ \citenamefont {Vargas}}]{frank04}%
  \BibitemOpen
  \bibfield  {author} {\bibinfo {author} {\bibfnamefont {A.}~\bibnamefont
  {Frank}}, \bibinfo {author} {\bibfnamefont {P.}~\bibnamefont {Van~Isacker}},
  \ and\ \bibinfo {author} {\bibfnamefont {C.~E.}\ \bibnamefont {Vargas}},\
  }\href {\doibase 10.1103/PhysRevC.69.034323} {\bibfield  {journal} {\bibinfo
  {journal} {Phys. Rev. C}\ }\textbf {\bibinfo {volume} {69}},\ \bibinfo
  {pages} {034323} (\bibinfo {year} {2004})}\BibitemShut {NoStop}%
\bibitem [{\citenamefont {Nomura}\ \emph {et~al.}(2010)\citenamefont {Nomura},
  \citenamefont {Shimizu},\ and\ \citenamefont {Otsuka}}]{nomura2010}%
  \BibitemOpen
  \bibfield  {author} {\bibinfo {author} {\bibfnamefont {K.}~\bibnamefont
  {Nomura}}, \bibinfo {author} {\bibfnamefont {N.}~\bibnamefont {Shimizu}}, \
  and\ \bibinfo {author} {\bibfnamefont {T.}~\bibnamefont {Otsuka}},\ }\href
  {\doibase 10.1103/PhysRevC.81.044307} {\bibfield  {journal} {\bibinfo
  {journal} {Phys. Rev. C}\ }\textbf {\bibinfo {volume} {81}},\ \bibinfo
  {pages} {044307} (\bibinfo {year} {2010})}\BibitemShut {NoStop}%
\bibitem [{\citenamefont {Nomura}\ \emph
  {et~al.}(2012{\natexlab{b}})\citenamefont {Nomura}, \citenamefont
  {Rodr\'iguez-Guzm\'an}, \citenamefont {Robledo},\ and\ \citenamefont
  {Shimizu}}]{nomura2012sc}%
  \BibitemOpen
  \bibfield  {author} {\bibinfo {author} {\bibfnamefont {K.}~\bibnamefont
  {Nomura}}, \bibinfo {author} {\bibfnamefont {R.}~\bibnamefont
  {Rodr\'iguez-Guzm\'an}}, \bibinfo {author} {\bibfnamefont {L.~M.}\
  \bibnamefont {Robledo}}, \ and\ \bibinfo {author} {\bibfnamefont
  {N.}~\bibnamefont {Shimizu}},\ }\href {\doibase 10.1103/PhysRevC.86.034322}
  {\bibfield  {journal} {\bibinfo  {journal} {Phys. Rev. C}\ }\textbf {\bibinfo
  {volume} {86}},\ \bibinfo {pages} {034322} (\bibinfo {year}
  {2012}{\natexlab{b}})}\BibitemShut {NoStop}%
\bibitem [{\citenamefont {Nomura}\ \emph {et~al.}(2013)\citenamefont {Nomura},
  \citenamefont {Rodr\'{\i}guez-Guzm\'an},\ and\ \citenamefont
  {Robledo}}]{nomura2013hg}%
  \BibitemOpen
  \bibfield  {author} {\bibinfo {author} {\bibfnamefont {K.}~\bibnamefont
  {Nomura}}, \bibinfo {author} {\bibfnamefont {R.}~\bibnamefont
  {Rodr\'{\i}guez-Guzm\'an}}, \ and\ \bibinfo {author} {\bibfnamefont {L.~M.}\
  \bibnamefont {Robledo}},\ }\href@noop {} {\bibfield  {journal} {\bibinfo
  {journal} {Phys. Rev. C}\ }\textbf {\bibinfo {volume} {87}},\ \bibinfo
  {pages} {064313} (\bibinfo {year} {2013})}\BibitemShut {NoStop}%
\bibitem [{\citenamefont {Bengtsson}\ \emph {et~al.}(1987)\citenamefont
  {Bengtsson}, \citenamefont {Bengtsson}, \citenamefont {Dudek}, \citenamefont
  {Leander}, \citenamefont {Nazarewicz},\ and\ \citenamefont
  {ye~Zhang}}]{bengtsson1987}%
  \BibitemOpen
  \bibfield  {author} {\bibinfo {author} {\bibfnamefont {R.}~\bibnamefont
  {Bengtsson}}, \bibinfo {author} {\bibfnamefont {T.}~\bibnamefont
  {Bengtsson}}, \bibinfo {author} {\bibfnamefont {J.}~\bibnamefont {Dudek}},
  \bibinfo {author} {\bibfnamefont {G.}~\bibnamefont {Leander}}, \bibinfo
  {author} {\bibfnamefont {W.}~\bibnamefont {Nazarewicz}}, \ and\ \bibinfo
  {author} {\bibfnamefont {J.}~\bibnamefont {ye~Zhang}},\ }\href {\doibase
  http://dx.doi.org/10.1016/0370-2693(87)91406-7} {\bibfield  {journal}
  {\bibinfo  {journal} {Physics Letters B}\ }\textbf {\bibinfo {volume}
  {183}},\ \bibinfo {pages} {1 } (\bibinfo {year} {1987})}\BibitemShut
  {NoStop}%
\bibitem [{\citenamefont {Bengtsson}\ and\ \citenamefont
  {Nazarewicz}(1989)}]{bengtsson1989}%
  \BibitemOpen
  \bibfield  {author} {\bibinfo {author} {\bibfnamefont {R.}~\bibnamefont
  {Bengtsson}}\ and\ \bibinfo {author} {\bibfnamefont {W.}~\bibnamefont
  {Nazarewicz}},\ }\href {\doibase 10.1007/BF01284554} {\bibfield  {journal}
  {\bibinfo  {journal} {Z. Phys. A}\ }\textbf {\bibinfo {volume} {334}},\
  \bibinfo {pages} {269} (\bibinfo {year} {1989})}\BibitemShut {NoStop}%
\bibitem [{\citenamefont {Nazarewicz}(1993)}]{nazarewicz1993}%
  \BibitemOpen
  \bibfield  {author} {\bibinfo {author} {\bibfnamefont {W.}~\bibnamefont
  {Nazarewicz}},\ }\href@noop {} {\bibfield  {journal} {\bibinfo  {journal}
  {Phys. Lett. B}\ }\textbf {\bibinfo {volume} {305}},\ \bibinfo {pages} {195 }
  (\bibinfo {year} {1993})}\BibitemShut {NoStop}%
\bibitem [{\citenamefont {{P. Van Isacker}}()}]{IBM1}%
  \BibitemOpen
  \bibfield  {author} {\bibinfo {author} {\bibnamefont {{P. Van Isacker}}},\
  }\href@noop {} {}\bibinfo {note} {Computer program IBM-1
  (unpublished)}\BibitemShut {NoStop}%
\bibitem [{\citenamefont {{Brookhaven National Nuclear Data Center}}()}]{data}%
  \BibitemOpen
  \bibfield  {author} {\bibinfo {author} {\bibnamefont {{Brookhaven National
  Nuclear Data Center}}},\ }\href@noop {} {}\bibinfo {howpublished}
  {{http://www.nndc.bnl.gov}}\BibitemShut {NoStop}%
\bibitem [{\citenamefont {Bai}\ \emph {et~al.}(2016)\citenamefont {Bai},
  \citenamefont {Li}, \citenamefont {L^^c3^^bc}, \citenamefont {Dong},
  \citenamefont {Wang},\ and\ \citenamefont {Zhang}}]{bai2016}%
  \BibitemOpen
  \bibfield  {author} {\bibinfo {author} {\bibfnamefont {H.-B.}\ \bibnamefont
  {Bai}}, \bibinfo {author} {\bibfnamefont {X.-W.}\ \bibnamefont {Li}},
  \bibinfo {author} {\bibfnamefont {L.-J.}\ \bibnamefont {L^^c3^^bc}}, \bibinfo
  {author} {\bibfnamefont {H.-F.}\ \bibnamefont {Dong}}, \bibinfo {author}
  {\bibfnamefont {Y.}~\bibnamefont {Wang}}, \ and\ \bibinfo {author}
  {\bibfnamefont {J.-F.}\ \bibnamefont {Zhang}},\ }\href
  {http://stacks.iop.org/1674-1137/40/i=7/a=074103} {\bibfield  {journal}
  {\bibinfo  {journal} {Chinese Physics C}\ }\textbf {\bibinfo {volume} {40}},\
  \bibinfo {pages} {074103} (\bibinfo {year} {2016})}\BibitemShut {NoStop}%
\bibitem [{\citenamefont {Kib\'edi}\ and\ \citenamefont
  {Spear}(2005)}]{kibedi2005}%
  \BibitemOpen
  \bibfield  {author} {\bibinfo {author} {\bibfnamefont {T.}~\bibnamefont
  {Kib\'edi}}\ and\ \bibinfo {author} {\bibfnamefont {R.}~\bibnamefont
  {Spear}},\ }\href {\doibase 10.1016/j.adt.2004.11.002} {\bibfield  {journal}
  {\bibinfo  {journal} {At. Data and Nucl. Data Tables}\ }\textbf {\bibinfo
  {volume} {89}},\ \bibinfo {pages} {77 } (\bibinfo {year} {2005})}\BibitemShut
  {NoStop}%
\end{thebibliography}%

\end{document}